\documentclass[twocolumn,preprint,authoryear,5p,times]{elsarticle}
\usepackage{amsmath}
\usepackage{natbib}
\usepackage[utf8]{inputenc}
\usepackage[T1]{fontenc}
\usepackage{pdfpages}
\usepackage{lipsum}
\usepackage{adjustbox}
\usepackage{aas_macros}
\usepackage[shortcuts]{extdash}
\usepackage{float}
\usepackage{bm}
\newcommand{\msun}{\,\mathrm{M}_\odot}

\DeclareUnicodeCharacter{00A0}{ }

\begin{document}

\title{Asteroid mass estimation using Markov-chain Monte Carlo}
\author[hki]{Lauri Siltala\corref{cor1}}
\ead{lauri.siltala@helsinki.fi}
\author[hki]{Mikael Granvik}
\ead{mgranvik@iki.fi}

\cortext[cor1]{Corresponding author}

\address[hki]{Department of Physics, P.O. Box 64, FI-00014 University
  of Helsinki, Finland}

\begin{abstract}
  Estimates for asteroid masses are based on their gravitational
  perturbations on the orbits of other objects such as Mars,
  spacecraft, or other asteroids and/or their satellites. In the case
  of asteroid-asteroid perturbations, this leads to an inverse problem
  in at least 13 dimensions where the aim is to derive the mass of the
  perturbing asteroid(s) and six orbital elements for both the
  perturbing asteroid(s) and the test asteroid(s) based on astrometric
  observations. We have developed and implemented three different mass
  estimation algorithms utilizing asteroid-asteroid perturbations: the
  very rough `marching' approximation, in which the asteroids' orbital
  elements are not fitted, thereby reducing the problem to a
  one-dimensional estimation of the mass, an implementation of the
  Nelder-Mead simplex method, and most significantly, a Markov-chain
  Monte Carlo (MCMC) approach. We describe each of these algorithms
  with particular focus on the MCMC algorithm, and present example
  results using both synthetic and real data. Our results agree with
  the published mass estimates, but suggest that the published
  uncertainties may be misleading as a consequence of using linearized
  mass-estimation methods. Finally, we discuss remaining challenges
  with the algorithms as well as future plans.
\end{abstract}


\maketitle

\section{Introduction}

As of March, 2017, over 700,000 minor planets have been discovered in
our solar system.  According to \citet{Car12} we have mass estimates
for 267 of these, half of which with an uncertainty of <20\% and 70\%
with an uncertainty of <50\%.  As such it is evident that we only know
the masses for a tiny fraction of all known asteroids, which shows
that a lot of work remains to be done in order to improve our
knowledge of their masses. Once an asteroid's mass and volume are
known, it is straightforward to compute its bulk density. The bulk
densities combined with measurements of the asteroids' surface
compositions help constrain their bulk compositions, information
which, in turn, is used to constrain the properties of the
protoplanetary disk from which the planets later formed as well as the
orbital evolution of the solar system \citep{2014Natur.505..629D}.

Gravitational perturbations by asteroids, the modeling of which
requires accurate mass estimates for them, currently represent the
greatest uncertainty in planetary ephemerides \citep{Sta00}. It is
relatively simple to determine a rough estimate for the mass of an
asteroid if its volume is known: one can simply use the mean bulk
density of the taxonomic class the asteroid belongs to and multiply
that by volume. The bulk density in turn can be estimated by using the
average bulk density of asteroids of the same spectral class.  This
does not however take into account things such as porosity, and the
bulk density of a given asteroid may certainly be quite different from
the mean of the taxonomic class. Thus, these rough estimates cannot be
considered very accurate.

For more useful and accurate mass estimations, one must study
gravitational perturbations caused by the selected asteroid upon
another body. Mars, other asteroids, satellites of the selected
asteroid and spacecraft have been used for this purpose
\citep{Hil02}. Mathematically the mass estimation procedure is an
inverse problem where the aim is to fit orbits for the considered
objects and masses for the perturbers to observational data
(i.e. astrometry).  Mass estimation is challenging, because individual
asteroids are quite small and thus have low masses, meaning that the
effect caused by their perturbations will also be very small. Close
asteroid encounters and precise astrometry of these encounters are
required for estimating asteroid masses based on perturbations on
other asteroids, which this work focuses on.

The first asteroid mass estimate was done by \citet{Her66}, where he
estimated a mass of $(1.17 \pm 0.1) \times 10^{-10}\msun$ for
(4)~Vesta based on its perturbations on (197)~Arete. This was done
with a least squares scheme, and {largely} agrees with the
modern {Dawn} estimate of {$(1.3025 \pm 0.0005) \times
  10^{-10} \msun$ \citep{Rus12}}. Within the next few decades other
mass estimations followed beginning with an estimate of $(6.7 \pm 0.4)
\times 10^{-10} \msun$ for (1)~Ceres using its perturbations on
(2)~Pallas by \citet{Sch70} which was also done with a least squares
scheme. {For reference, the modern Dawn estimate for the mass
  of (1)~Ceres is $(4.7179 \pm 0.0005) \times 10^{-10} \msun$
  \citep{Rus16}.} {A few years afterwards, (2)~Pallas’s mass
  was determined to be $(1.3 \pm 0.4) \times 10^{-10} \msun$ by
  \citet{Sch74} from its perturbations on (1)~Ceres.} The fourth
asteroid to have its mass estimated was (10)~Hygiea using
(829)~Academia as its test asteroid by \citet{Sch87} with a least
squares fit. The determined value was $(4.7 \pm 2.3) \times 10^{-11}
\msun$. Up until this point, all mass estimations had been performed
using asteroid-asteroid perturbations. This changed when \citet{Sta89}
determined the masses of (1)~Ceres, (2)~Pallas, and (4)~Vesta from
their perturbations on Mars utilizing data taken by the Viking
lander. The study resulted in masses of $(5.0 \pm 0.2) \times 10^{-10}
\msun$, $(1.4 \pm 0.2) \times 10^{-10} \msun$ and $(1.5 \pm 0.3)
\times 10^{-10} \msun$ for the three asteroids, respectively, which
was in line with earlier results. Eight years later, a third mass
determination method surfaced when \citet{Pet97} determined the mass
of (243)~Ida based on the trajectory of its satellite Dactyl as
observed by the Galileo spacecraft. Based on the assumption that
Dactyl's orbit is stable, they determined a value of $(2.2 \pm 0.3)
\times 10^{-14} \msun$ for the mass of (243)~Ida. During the same year
a fourth method was introduced when \citet{Yeo97} inferred
(253)~Mathilde's mass based on its perturbations on the NEAR
spacecraft during a close encounter. The resulting mass of
(253)~Mathilde was $(5.19 \pm 0.02) \times 10^{-14} \msun$. Note, in
particular, that the formal uncertainty of this estimate is
substantially smaller compared to earlier estimates. Due to the high
accuracy of the Doppler shift of the radio-communications signal, this
is the most accurate mass estimation method. In the near future, the
very accurate astrometry produced by the Gaia mission \citep{Pru16}
will allow the estimation of masses with a relative precision better
than 50\% for about 100 asteroids \citep{Mou08}.

As the first asteroid mass estimation was done half a century ago, the
principles behind asteroid mass estimation based on asteroid-asteroid
close encounters are hardly new. However, previous estimates have
largely relied on linearized least-squares methods which may cause
issues with uncertainties of the mass estimates in particular. This
issue is illustrated by Fig.~2 in \citet{Car12}, which shows that that
there is significant disagreement between the uncertainties of
different independent mass estimates for asteroid (52)~Europa, which
suggests that the uncertainties may be too low. This may be a
consequence of the used linearized least-squares methods which provide
uncertainty information based on certain assumptions, such as Gaussian
distributions.

To this end, we have developed new asteroid mass estimation methods
based on asteroid-asteroid close encounters and implemented these in
the OpenOrb software \citep{Gra09}. Our main focus is a Markov-chain
Monte Carlo (MCMC) algorithm, but we also introduce algorithms based
on the Nelder-Mead simplex algorithm and the so-called marching
approximation, which reduces the problem to one dimension.

MCMC directly provides us with probability distributions of each model
parameter without there being any need to assume any specific
underlying distribution and thus we should obtain more accurate
uncertainties.  This is one of the main advantages we perceive our
MCMC approach to have in comparison to the earlier mass estimations
algorithms.

\section{Theory and methods}

The case of asteroid-asteroid perturbations leads to a
multi-dimensional inverse problem where the aim is to derive the mass
of the perturbing asteroid and six orbital elements at a specific
epoch for both the perturbing asteroid(s) and the test asteroid(s)
using astrometric observations taken over a relatively long
timespan. A single astrometric observation consists of measurements of
Right Ascension and Declination at a specific point in time.

{The $\chi^2$ test statistic represents the goodness of fit
  resulting from a set of model parameters $\bm{P}$:}
\begin{equation}
  \chi^2 = \sum_{i=1}^{N_\mathrm{obj}} \sum_{j=1}^{N_{\mathrm{obs},i}}
  \left[ \frac{(\alpha_{i,j}^0 -
      \alpha_{i,j}(\bm{P}))\cos{\delta_{i,j}^0}}{\sigma_{\alpha,i,j}}
    - \frac{\delta_{i,j}^0 -
      \delta_{i,j}(\bm{P})}{\sigma_{\delta,i,j}}\right]
\end{equation}
{where $N_\mathrm{obj}$ is the number of asteroids included,
  $N_{\mathrm{obs},i}$ is the number of observations used for asteroid
  $i$, $\alpha_{i,j}^0$ and $\delta_{i,j}^0$ are the observed Right
  Ascension (RA) and Declination (Dec), respectively, of asteroid $i$
  at time $t_j$, $\alpha_{i,j}(\bm{P})$ and $\delta_{i,j}(\bm{P})$ are
  the predicted RA and Dec of asteroid $i$ at time $t_j$, and
  $\sigma_{\alpha,i,j}$ and $\sigma_{\delta,i,j}$ are the standard
  deviations of the astrometric uncertainties corresponding to the
  observations with the same indices. The closer the predicted
  positions are to the observations, the lower the $\chi^2$ statistic
  is, and thus a smaller value corresponds to a better agreement
  between observations and model prediction.}

To describe the results we use the reduced $\chi^2$ test statistic:
\begin{equation}
\chi^2_\mathrm{red} = \frac{\chi^2}{K}
\end{equation}
{where $K$ denotes the degrees of freedom in the model, that is,}
\begin{equation}
K = 2 \sum_i N_\mathrm{obs,i} - 6N_\mathrm{obj} - N_\mathrm{per}
\end{equation}
{or the total amount of observations minus the total amount of
  fitted parameters, that is, six orbital elements for each asteroid
  and the masses of the perturbing asteroids. $\chi^2_\mathrm{red}=1$
  implies a good fit whereas $\chi^2_\mathrm{red}>1$ imply poor fits
  and $\chi^2_\mathrm{red}<1$ suggests that the model parameters
  (partly) describe the noise rather than the physical model only. The
  statistic is sensitive to the assumed astrometric uncertainties
  which, considering our current lack of a proper observational error
  model, may lead to suboptimal values except in the case of synthetic
  astrometry where we know the uncertainty of each data point. The
  sensitivity to assumptions is particularly noteworthy when
  considering the results of the mass marching algorithm, because we
  currently base our estimates for the observational errors on the RMS
  values corresponding to the best-fit solution produced by the
  marching algorithm.}

To compute $\alpha_{i,j}(\bm{P})$ and $\delta_{i,j}(\bm{P})$ we
integrate the orbits of the asteroids through the observational
timespan while taking the gravitational perturbations of both the
perturbing asteroid and the planets into account. We parameterize the
orbits with heliocentric osculating Cartesian state vectors at a
specific epoch $t$, that is, $\bm{S} =
(x,y,z,\dot{x},\dot{y},\dot{z})$ where $(x,y,z)$ is the asteroid's
position and $(\dot{x},\dot{y},\dot{z})$ its velocity at epoch $t$. In
general, the total set of parameters is thus
$\bm{P}=(\bm{S}_1,\bm{S}_2,\ldots,\bm{S}_{N_\mathrm{obj}},M_1,M_2,\ldots,M_{N_\mathrm{per}})$,
where $M_i$ is the mass of the $i$th perturber and $N_\mathrm{per}$ is
the number of perturbers considered. In what follows we only consider
cases with $N_\mathrm{obj}=2$ and $N_\mathrm{per}=1$. {In this
  case, as no asteroids are perturbing the perturber itself the
  initial perturber orbit can be assumed to not change
  significantly. Despite this, it still needs to be included in the
  model as there is always some degree of uncertainty in the
  perturbing orbit as well. Fitting for the perturber's orbital
  elements will automatically account for these uncertainties, and
  affect the resulting mass as the gravitational perturbations on the
  test asteroid depend on the perturber's orbit in addition to the
  mass.}

Our mass marching and Nelder-Mead algorithms both seek to minimize the
sum of the $\chi^2$ values for both used asteroids in order to find
the best possible fit whereas the Markov-chain Monte Carlo method
samples a region of possible model parameters in order to determine
the probability distributions of each model parameter. In the
following sections, we individually describe each of these algorithms.

\subsection{The mass marching approximation}

The first, and simplest by far, mass estimation algorithm is the mass
marching approximation.  In this method, an intitial mass estimate is
computed from observed H-magnitudes of the perturbing asteroid
assuming a spherical shape for the asteroid \citep{Che04}:
\begin{equation}
  M_\mathrm{init} = \frac{\pi}{6}\rho D^3\,,
\end{equation}
where we assume a value of 2.5 g/cm$^3$ for the density $\rho$. The
diameter $D$ is \citep{Che04}:
\begin{equation}
  D = 1329 \times 10^{-H/5}\,p_V^{-1/2}\,\mathrm{km}\,,
\end{equation}
where we assume a value of 0.15 for the geometric albedo $p_V$.  A
range of masses around this estimate are then `marched'
through. Specifically, the masses range from $0.2M_\mathrm{init}$ to
$3.0 M_\mathrm{init}$ at intervals of $0.01 M_\mathrm{init}$. For each
mass we compute the corresponding $\chi^2$ and we thus obtain the
dependence of $\chi^2$ on perturber mass. The initial orbital elements
are considered constant, that is, the problem is reduced from 13
dimensions to one dimension. The orbits do however change outside the
initial epoch as a result of perturbations applied during the
integration process.  The approximations allow fast execution
typically requiring some minutes for the entire analysis.

As the initial orbital elements { are not fitted in the marching
  algorithm but remain constant}, this is a rough approximation that
does not give as accurate results as the other two methods. { We
  note that} the initial orbits are usually suboptimal, both because
of orbital uncertainties and because the initial orbits we use do not
account for perturbations by the perturbing asteroid. The method is
nevertheless useful for rough estimates due to its simplicity and
speed, and in some cases even delivers surprisingly good
results. Another potential use for the algorithm is determining
whether mass estimation is possible for a given close encounter to
begin with: if the tested pair does not have any useful close
encounters, mass will not affect the $\chi^2$ values.  This method is
also employed by the MCMC and simplex methods to determine starting
values for asteroid masses where possible.

\subsection{The Nelder-Mead simplex method}

The Nelder-Mead method \citep{Nel65} is based on a simplex, which
refers to a geometric object consisting of $n+1$ vertices in a
$n$-dimensional space. In our case, we are dealing with 13 separate
variables consisting of the six orbital elements for both the
perturber and test asteroid and the mass of the perturber, that is,
the vector $\bm{P}$.  Thus we are dealing with a 13 dimensional space,
and 14 vertices are required. These vertices are perhaps easiest to
understand as unique sets of values for each variable; in other words,
each vertex consists of orbits for both asteroids and the perturber
mass, each slightly different in comparison to other vertices.

The starting values for each vertex are created as follows: the first
vertex directly uses the given input state vectors for the asteroids
--- for example, those determined by OpenOrb's least-squares
method. The elements of the other vertices are created such that for a
given vertex $n$, the elements used are computed according to the
formula
$(1+\epsilon)^{n-8}(\bm{S}_{1,\mathrm{in}},\bm{S}_{2,\mathrm{in}})$,
where $(\bm{S}_{1,\mathrm{in}},\bm{S}_{2,\mathrm{in}})$ refers to the
input state vectors. For $\epsilon$ we currently use a value of $3
\times 10^{-7}$.  As $n$ ranges from 2 to 14, the exponents will be
$-6, -5, -4 {\ldots} 6$ and it is ensured that vertices will be
distributed on both sides of the input orbit.

In regards to masses, the first vertex's mass is estimated using the
marching algorithm where possible; the code will run the marching
algorithm automatically, and if it produces a non-zero best-fit mass,
this is used as the initial mass. If the marching algorithm fails,
i.e., provides a mass of zero, the initial mass used for the marching
algorithm itself is used. At this stage, we also apply our outlier
rejection algorithm to the observations. This algorithm rejects all
data points with residuals $>4\sigma$. Furthermore, we also set
observational errors separately for each asteroid based on root mean
square, or RMS, values for the observations based on the best fit of
the marching algorithm.  With this done, we derive the other vertices
from the initial mass much like we did for the orbital elements; the
only difference is a significantly higher $\epsilon$ value of $5
\times 10^{-2}$, because it is assumed that the initial mass is not
very accurate.  Once the starting vertices have been prepared and the
marching algorithm has been {run}, the Nelder-Mead algorithm
can begin.  The algorithm continuously updates the simplex's vertices
toward a better solution, eventually converging to the best one. For a
more detailed discussion of the Nelder-Mead algorithm itself, we refer
the interested reader to the original paper \citep{Nel65}.

\subsection{MCMC mass estimation} %


The general idea of Markov-chain Monte Carlo, or MCMC, algorithms is
to create a Markov chain to estimate the unknown posterior probability
distributions of the parameters $p(\bm{P})$ of a given model. A Markov
chain, in turn, is a construct consisting of a series of elements in
which each element is derived from the one preceding it. In a properly
constructed Markov chain the posterior distributions of individual
elements in the chain match the probability distributions of these
elements. Thus as the end result of MCMC, one gets the probability
distributions of each parameter in the model. From these
distributions, one can directly determine the maximum-likelihood
values from the peaks of the distributions alongside the confidence
limits. These confidence limits are the main advantage we see in MCMC
for the mass estimation problem. The limits should be quite accurate,
as we do not need to make any assumptions regarding the shape of the
posterior distribution.  As mentioned in the introduction, it is
common to assume a Gaussian shape for the posterior distribution, but
as our results will show, the posterior probability distribution of
the mass is often non-Gaussian.

As in the case with the Nelder-Mead algorithm described above, we
begin our MCMC algorithm by running the marching algorithm on the
data, rejecting outliers and setting observational errors in the same
manner. Once this is done, the MCMC chain itself begins, starting with
the initial mass determined exactly like in the Nelder-Mead case,
while the initial orbits are again those previously calculated by,
e.g., the least-squares method and given as part of the input data.

Our MCMC method is based on the Adaptive Metropolis (AM) algorithm
\citep{Haa02}. Proposed parameters $\bm{P}'$ are generated by adding
deviates $\bm{\Delta P}$ to the previously accepted, or $i$th, set of
parameters $\bm{P}_i$:
\begin{equation}
    \bm{P}' = \bm{P}_i + \bm{\Delta P}\,.
\end{equation}
The deviates are computed as
\begin{equation}
  \bm{\Delta P} = \bm{A} \bm{R}\,,
\end{equation}
where $\bm{A}$ is the Cholesky decomposition of the proposal
distribution $\bm{S_i}$ and $\bm{R}$ is a
$(6N_\mathrm{obj}+N_\mathrm{per})$-vector (here 13-vector) consisting
of Gaussian-distributed random numbers. {At this point,
  proposals with negative masses are automatically rejected as they
  are not physically plausible while all masses greater than or equal
  to zero are permitted.}

The proposal distribution $\bm{S_i}$ is described by the parameters'
covariance matrix, and it is constantly updated based on the computed
chain itself using the empirical covariance matrix formula
\citep{Haa02}:
\begin{equation}
  \bm{S_i} = \frac{2.4^2}{d} \frac{1}{i - 1} \sum_{j=1}^i (\bm{P_j} -
  \overline{\bm{P}})(\bm{P_j} - \overline{\bm{P}})^{\rm{T}} + \epsilon
  \bm{I_d}\,,
\end{equation}
where $2.4^2/d$ is a scaling parameter which has been shown to
optimize the chain for Gaussian distributions, $d$ represents the
{number} of dimensions in the model (and thus here $d=13$),
$\bm{P}_j$ represents all of the accepted solutions in the chain,
$\overline{\bm{P}}$ represents their mean, $\bm{I_d}$ is the identity
matrix, and $\epsilon$ is an arbitrary small parameter. We empirically
found that $\epsilon=10^{-26}$ produces good results and that the
results are not particularly sensitive to its value. Posterior
probability density ($p$) is then obtained as
\begin{equation}
  p(\bm{P}') \propto \exp(-\frac{1}{2} \chi^2(\bm{P}')) \,.
\end{equation}

Next, the posterior probability density is compared to the previously
accepted solution:
\begin{equation}
  a_r = \frac{p(\bm{P}')}{p(\bm{P}_i)} = \exp(-\frac{1}{2} \left(\chi^2(\bm{P}') - \chi^2(\bm{P})\right))
\end{equation}
If $a_r > 1$, the proposed solution is better than the previously
accepted solution and hence it is automatically accepted as the next
transition. Otherwise, it is accepted with a probability of $a_r$. The
first proposal in the chain is always accepted, because there is no
previous solution to compare with. Should the proposal be accepted and
the chain length $i >= 19$, we also update the covariance matrix as
described above.

We repeat the process until the desired amount of transitions is
reached. We have typically required 50,000 transitions.  Halfway
through the run, a new chain is started with an initial mass of
$2M_\mathrm{init}$ and the same orbital elements as used to initiate
the first chain. This is done both to ensure that a sufficiently large
range of masses is tested, and to ensure that the parameters converge
to the same posterior distribution with different starting values.

{We determine our confidence limits by calculating a
  kernel-density estimate (KDE) based on the statistics of repetitions
  such that the limits encompassing 68.26\% of the probability mass
  around the peak of the KDE correspond to $1\sigma$ while the limits
  encompassing 99.73\% of the probability mass correspond to
  $3\sigma$.}

Initially we used a more standard Metropolis-Hastings (MH) algorithm,
but encountered convergence and mixing issues in many cases with
several model parameters. We suspected that the cause might have been
issues in the initial covariance matrices that we obtained using
separate least-squares solutions for both asteroids. AM in general is
designed to correct such issues and in practise it exceeded our
expectations: AM provided covariance matrices differing from our
initial matrices and our issues completely disappeared with the new
adapted matrices.  In addition to this, AM permits us to use only a
single covariance matrix with all model parameters included for the
chain.  With MH, we used covariance matrices calculated separately for
each object with the least-squares method. These matrices did not take
perturbations into account, and thus we did not account for
correlations between orbital elements for different asteroids or
between orbital elements and the mass of the perturber. These
limitations are no longer present when using AM as we only rely on the
initial block-matrix at the beginning of each run and subsequently
update it with the, typically, non-zero correlations. For a more
technical and in-depth review of MCMC algorithms we refer the
interested reader to \citet{Fei12} and references therein.

\section{The data}

We selected nine different encounters, all of which were included in
the recent work of \citet{Bae11}, allowing us to directly compare our
results to theirs in addition to the values of \citet{Car12}, which
are weighted averages of all mass estimates available for each
object. The selected encounters and their epochs are shown in Table
\ref{astpairs}. In our notation, we separate the numbers or
designations of perturbing asteroids and the massless test asteroids
with a semicolon, e.g., [7;17186] means that (7)~Iris is the
perturbing asteroid and (17186)~Sergivanov the test
asteroid. {This abbreviated notation is used so as to be
  forwards compatible with future papers involving larger amounts of
  perturbing and/or test asteroids.} One can see that all of these
encounters happened after the year 1990; this is a deliberate choice
taken because earlier astrometry is more inaccurate in comparison to
more modern astrometry.  Our algorithm currently lacks a proper
observational error model, which would have a significant impact if
older data was used. {This leads to the test cases of
  {[29;987]} and {[52;124]} in particular to have a very limited
  amount of astrometry from before the close encounter epoch, which
  explains the large uncertainties.}

We used all of the astrometry available through the Minor Planet
Center (MPC) for the selected asteroids between January 1990 and March
2016.  We obtained initial orbits from the MPCORB
database. {Table \ref{errs} shows the root-mean-square (RMS)
  values of the residuals and the number of outliers for each object
  resulting from a least-squares fit of the orbital elements only.}

\begin{table}
  \caption{The encounters used in this work. The numbers in the first
    column represent the numbers of used asteroids.  The first
    asteroid represents the perturber while the second represents the
    test asteroid.  Dates of close encounters taken from \protect
    \citet{Gal02} with the exceptions of [7;17186] and [15;14401],
    which are taken from their
    website\protect\footnote{http://staryweb.fmph.uniba.sk/index.php?id=2171},
    and [19;27799] for which we estimated a date ourselves.  Note that
    reference mass 1 is based on a single estimate \protect
    \citep{Bae11} whereas reference mass 2 is the weighted average of
    many mass determinations \protect \citep{Car12}.}
  \label{astpairs}
  \begin{center}
    \begin{tabular}{cccc}
      \hline
      Encounter	 & Date of CE & Ref.~mass.~1     & Ref.~mass.~2 \\
      & 							& [$10^{-11} \msun$] & [$10^{-11} \msun$] \\
      \hline
	  {[7;17186]}  & 1998/3/14    & $0.534  \pm 0.075$   & $0.649 \pm 0.106$              \\ 
	  {[10;3946]}  & 1998/3/30    & $4.051 \pm 0.1$      & $4.34 \pm 0.26$                \\ 
	  {[13;14689]} & 1997/7/21    & $0.8   \pm 0.22$     & $0.444 \pm 0.214$              \\ 
	  {[15;14401]} & 2005/7/15    & $1.427 \pm 0.14$	   & $1.58 \pm 0.09$ 						   \\ 
	  {[19;3486]}  & 1996/5/14    & $0.39  \pm 0.037$	   & $0.433 \pm 0.073$						     \\ 
	  {[19;27799]} & 1997/12/24   & $0.91  \pm 0.16$     & $0.433 \pm 0.073$					     \\ 
	  {[29;987]}   & 1994/3/3     & $0.773 \pm 0.032$		 & $0.649 \pm 0.101$					     \\ 
	  {[52;124]} 	 & 1993/10/17   & $1.139 \pm 0.079$		 & $1.20 \pm 0.29$ 						   \\ 
	  {[704;7461]} & 1997/5/31    & $1.97  \pm 0.59 $		 & $1.65 \pm 0.23$ \\
      \hline
    \end{tabular}
  \end{center}
\end{table}

\begin{table*}
  \caption{{RMS values for right ascension and declination
      residuals (in arcseconds) and total numbers of observations and
      outliers for each object. RMS$_1$ corresponds to data prior to
      outlier removal and RMS$_2$ in turn to data with outliers
      removed.}}
  \label{errs}
  \begin{center}
    \begin{tabular}{lcccccc}
      \hline
      Object	          & RMS$_1$(Dec) & RMS$_1$(RA)     & N$_\mathrm{outlier}$  & N$_\mathrm{total}$ & RMS$_2$(Dec) & RMS$_2$(RA)\\
      \hline
	  (7) Iris            & 0.83         & 0.99            & 46                & 2792           & 0.60         & 0.63       \\
	  (10) Hygiea         & 0.51         & 0.67            & 80                & 3640           & 0.43         & 0.51       \\
	  (13) Egeria         & 0.55         & 0.56            & 48                & 2600           & 0.41         & 0.50       \\
      (15) Eunomia        & 1.00         & 1.30            & 22                & 2726           & 0.57         & 0.48       \\
      (19) Fortuna        & 0.55         & 0.68            & 68                & 3182           & 0.46         & 0.57       \\
      (29) Amphitrite     & 0.74         & 0.78            & 30                & 2538           & 0.46         & 0.54       \\
      (52) Europa         & 0.63         & 0.65            & 54                & 3128           & 0.53         & 0.55       \\
      (124) Alkeste       & 0.67         & 1.04            & 52                & 3454           & 0.51         & 0.62       \\
      (704) Interamnia    & 0.50         & 0.49            & 110               & 4334           & 0.40         & 0.38       \\
      (987) Wallia        & 0.56         & 0.48            & 38                & 3058           & 0.51         & 0.44       \\
      (3486) Fulchignoni  & 0.58         & 0.62            & 32                & 2098           & 0.53         & 0.57       \\
	  (3946) Shor         & 0.73         & 0.54            & 32                & 4224           & 0.47         & 0.49       \\
	  (7461) Kachmokiam   & 0.51         & 0.59            & 30                & 2382           & 0.48         & 0.54       \\
	  (14401) Reikoyukawa & 0.63         & 0.66            & 16                & 2140           & 0.54         & 0.64       \\
	  (14689) 2000 AM$_2$ & 0.55         & 0.60            & 18                & 2360           & 0.51         & 0.58       \\
	  (17186) Sergivanov  & 0.55         & 0.61            & 12                & 1628           & 0.51         & 0.60       \\
	  (27799) 1993 FQ$_{23}$ & 0.63         & 0.71            & 10                & 1264           & 0.61         & 0.65       \\

      \hline
    \end{tabular}
  \end{center}
\end{table*}
Besides the real asteroids and observations, we also generated a
synthetic test case using OpenOrb with a perturber mass of $8.852
\times 10^{-11}\msun$, where $\msun$ refers to the mass of the Sun,
and no perturbations apart from the single perturbing asteroid. For
the test asteroid and the perturber, we generated synthetic astrometry
and added noise using standard deviations of 0.05" and 0.01",
respectively. Residuals similar to this are to be expected for
correctly working code.  Synthetic astrometry has major advantages for
testing mass estimation algorithms: we know the exact mass of the
perturber, planetary perturbations can be ignored as the data was
generated without them, greatly shortening the computation time, and
perturbations by unmodeled asteroids do not exist.

\section{Results and discussion}

\subsection{Results for mass marching}

\begin{figure}
  \begin{center}
    \includegraphics[width=1.0\columnwidth]{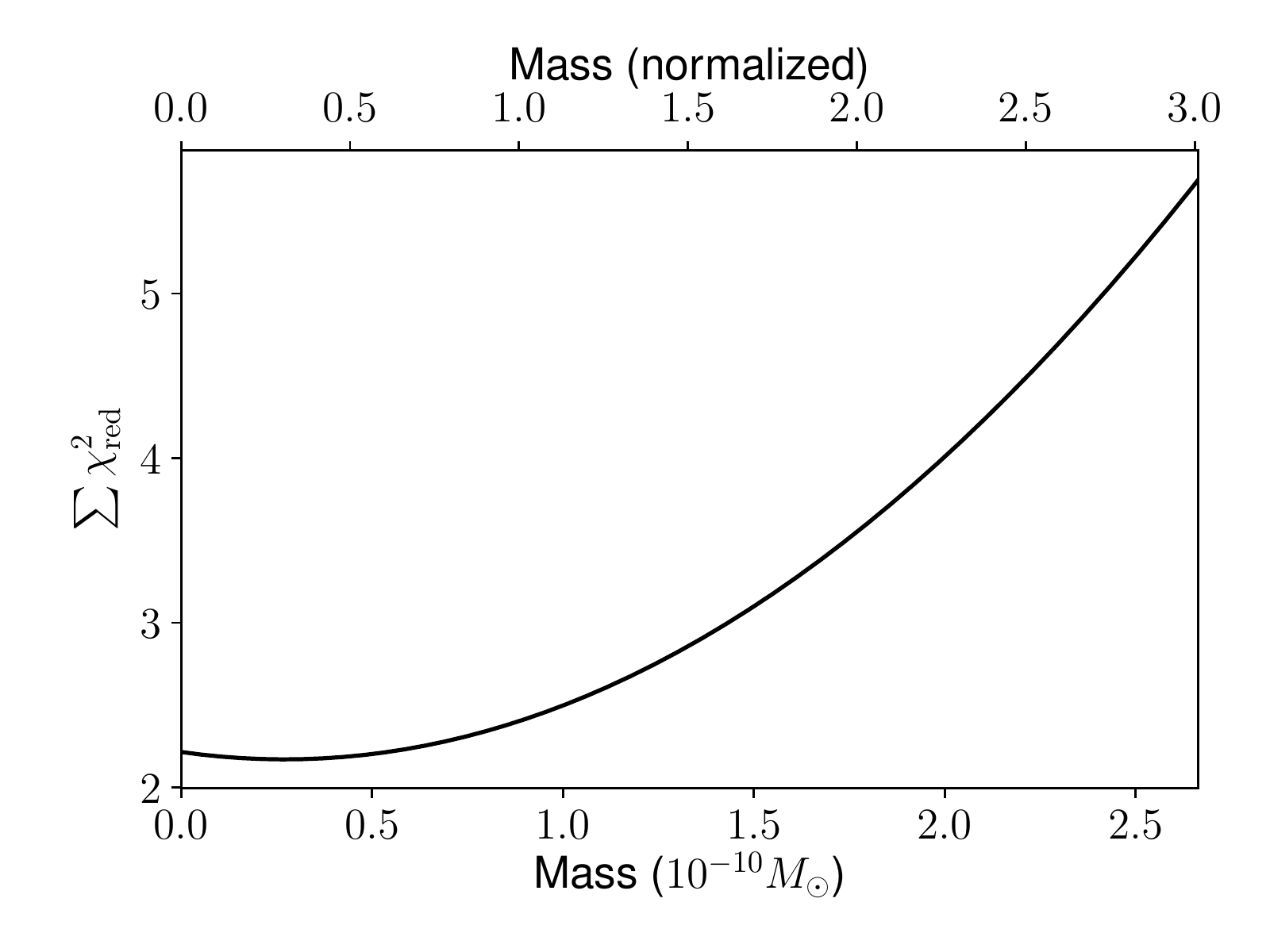}
    \caption{Results of the mass marching algorithm applied to
      synthetic data. The sum of the $\chi^2_\mathrm{red}$ values for
      both asteroids is shown on the $y$-axis. The bottom $x$-axis
      represents perturber mass in solar mass $\msun$, while the top
      $x$-axis is normalized such that 1 represents the correct mass,
      2 is twice that, etc.}
    \label{synth_march}
  \end{center}
\end{figure}

The results of the mass marching algorithm for the synthetic case are
shown in Fig.~\ref{synth_march}. The orbits were obtained with
OpenOrb's least-squares algorithm which does not take asteroidal
perturbations into account. This is expected to give worse results due
to { unperturbed} initial orbits, but is done nonetheless for the
case to be more comparable to real data, where we do not have {
  fully-perturbed} initial orbits.  The best mass, i.e., the one
resulting in the lowest total $\chi^2 = 1144$, is $2.744 \times
10^{-11}\msun$. A total of 540 observations was used for this case,
which leads to a reduced $\chi^2_\mathrm{red} = 2.17$, signifying a
decent fit. The best fitted mass is roughly 30\% of the correct mass
of $8.852 \times 10^{-11}\msun$, and from the figure one can see that
for these orbits, the correct mass results in significantly worse
fits. Even so, in this case the marching algorithm clearly detects the
existence of the gravitational perturbation.  The general shape of the
curve also looks very reasonable; it is intuitive that a single
minimum would be found and that the fit becomes worse with greater
offset from the best mass.

\begin{figure}
  \begin{center}
    \includegraphics[width=1.0\columnwidth]{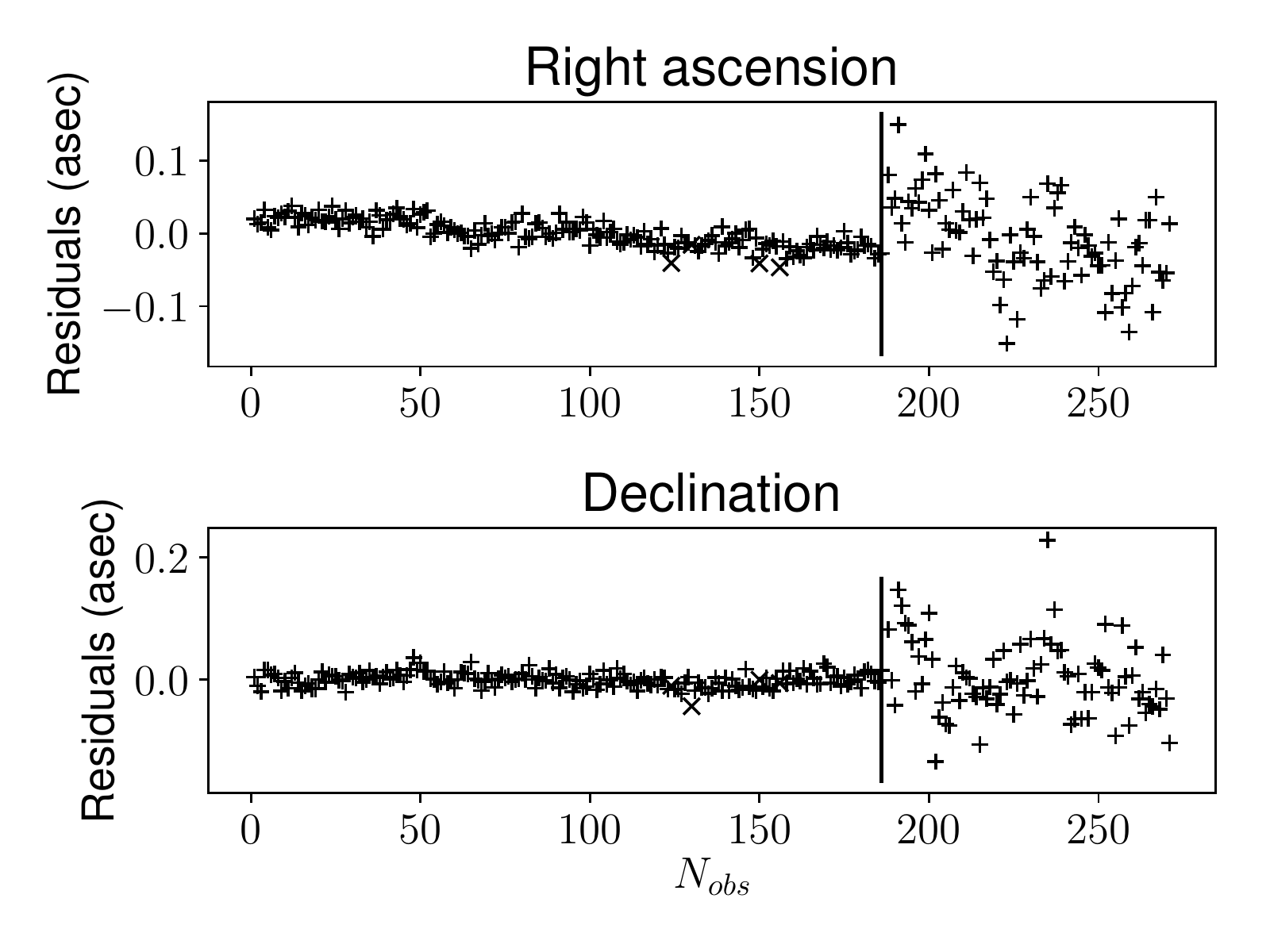}
    \caption{Residuals of the best fit for synthetic data obtained
      with the mass marching method. The data for the perturber and
      the test asteroid are separated by the vertical line, with the
      perturber's data on the left side. The data points are sorted by
      observation time in such a manner that each asteroid's data
      begins with the earliest observation and ends with the last.}
    \label{synth_march_residuals}
  \end{center}
\end{figure}

Residuals corresponding to the best-fit solution are shown in
Fig.~\ref{synth_march_residuals}. They are very small, which is
expected, since the synthetic data was generated with very small
errors. Residuals are noticeably higher for the test asteroid than for
the perturber, which is expected as the test asteroid's astrometry had
errors approximately 5 times larger than those of the
perturber. Interestingly, there appears to be a slight linear trend in
the right ascension residuals, where they appear to slightly decrease
with time in a linear trend. This is caused by a too small semi-major
axis leading to a too large mean motion and, further, systematic
decrease of the $O-C$ residual in RA.

\begin{figure}
  \begin{center}
    \includegraphics[width=1.0\columnwidth]{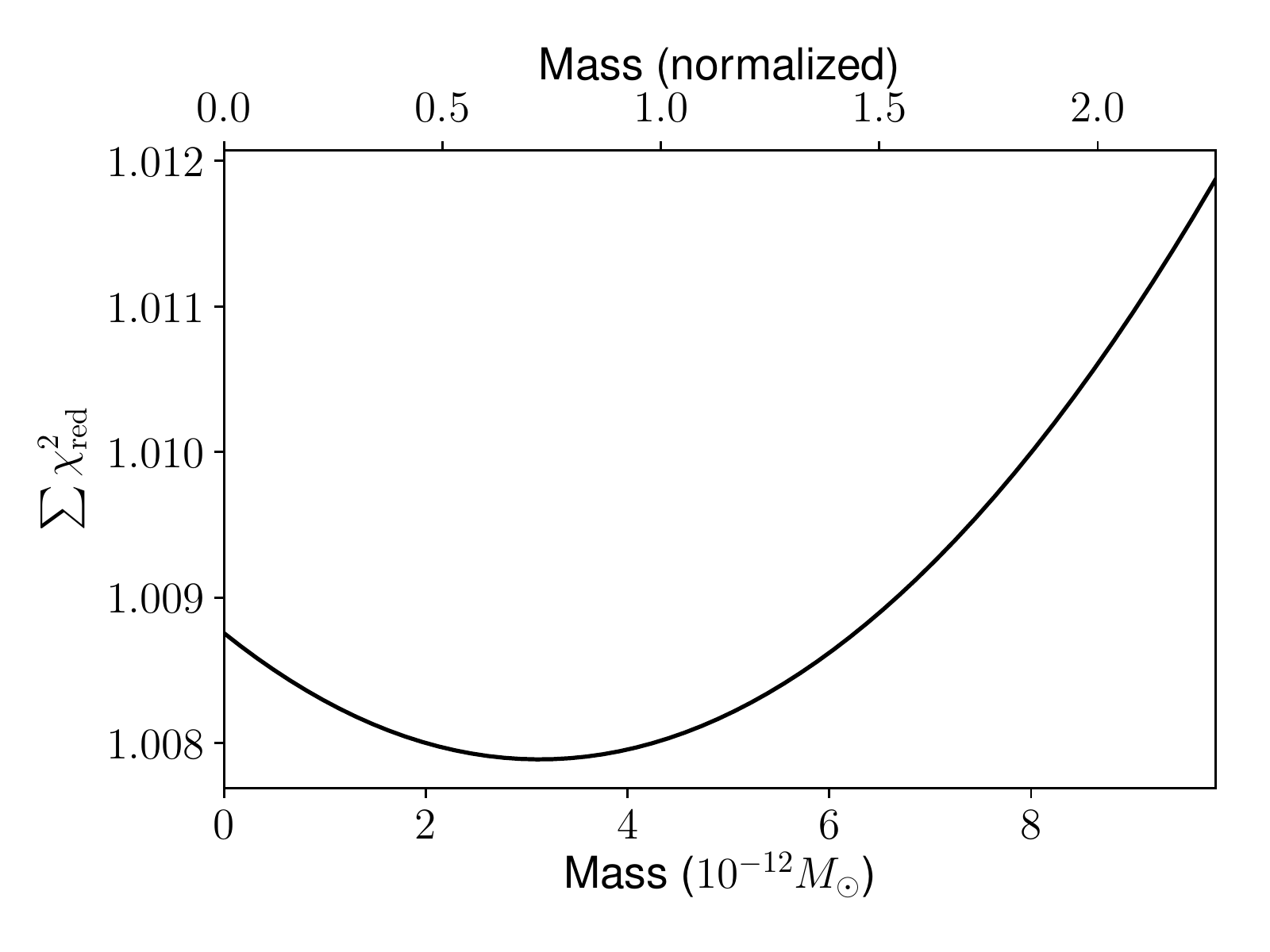}
    \caption{Results of the mass marching algorithm applied to the
      [19;3486] encounter. The sum of the $\chi^2_\mathrm{red}$ values
      for both asteroids is shown on the $y$-axis. The bottom $x$-axis
      represents perturber mass in solar mass $\msun$, while the top
      $x$-axis is normalized such that 1 represents the mass from
      \citet{Car12}, 2 is twice that, etc.}
    \label{19_march}
  \end{center}
\end{figure}

We also ran the marching algorithm for all real encounters
considered. Similarly to the synthetic case, we determined the initial
orbits with least-squares. In the case of [19;3486], the marching
algorithm finds a non-zero best fit mass that is approximately 75\% of
the mass from \citet{Car12}, which is actually a very good result
considering the approximations used (Fig.~\ref{19_march}).

\begin{figure}
  \hspace*{-10cm}
  \begin{center}
    \includegraphics[width=1.0\columnwidth]{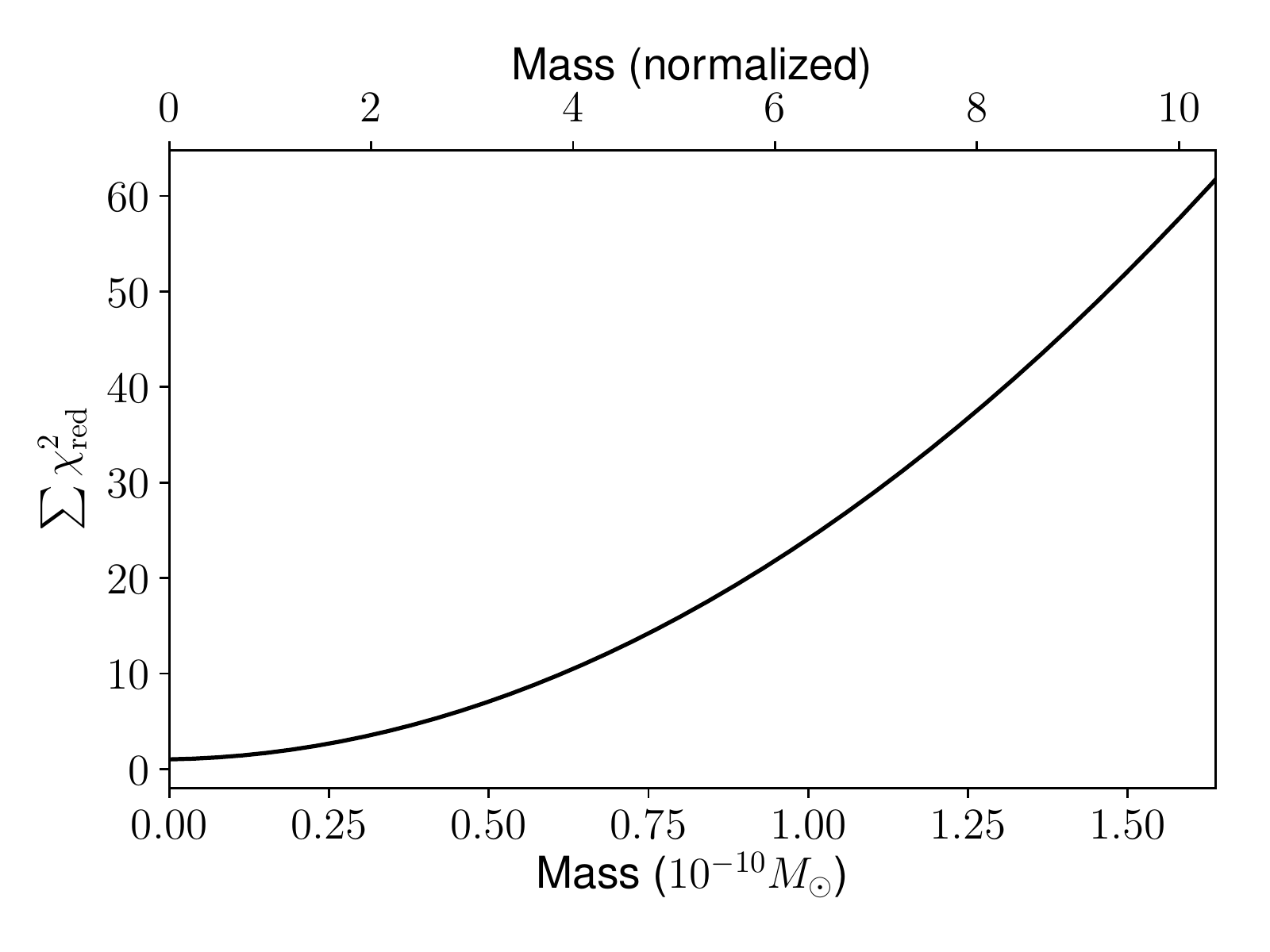}
    \caption{Results of the mass marching algorithm applied to the
      [15;14401] encounter.}
    \label{15_march}
  \end{center}
\end{figure}

\begin{figure}
  \begin{center}
    \includegraphics[width=1.0\columnwidth]{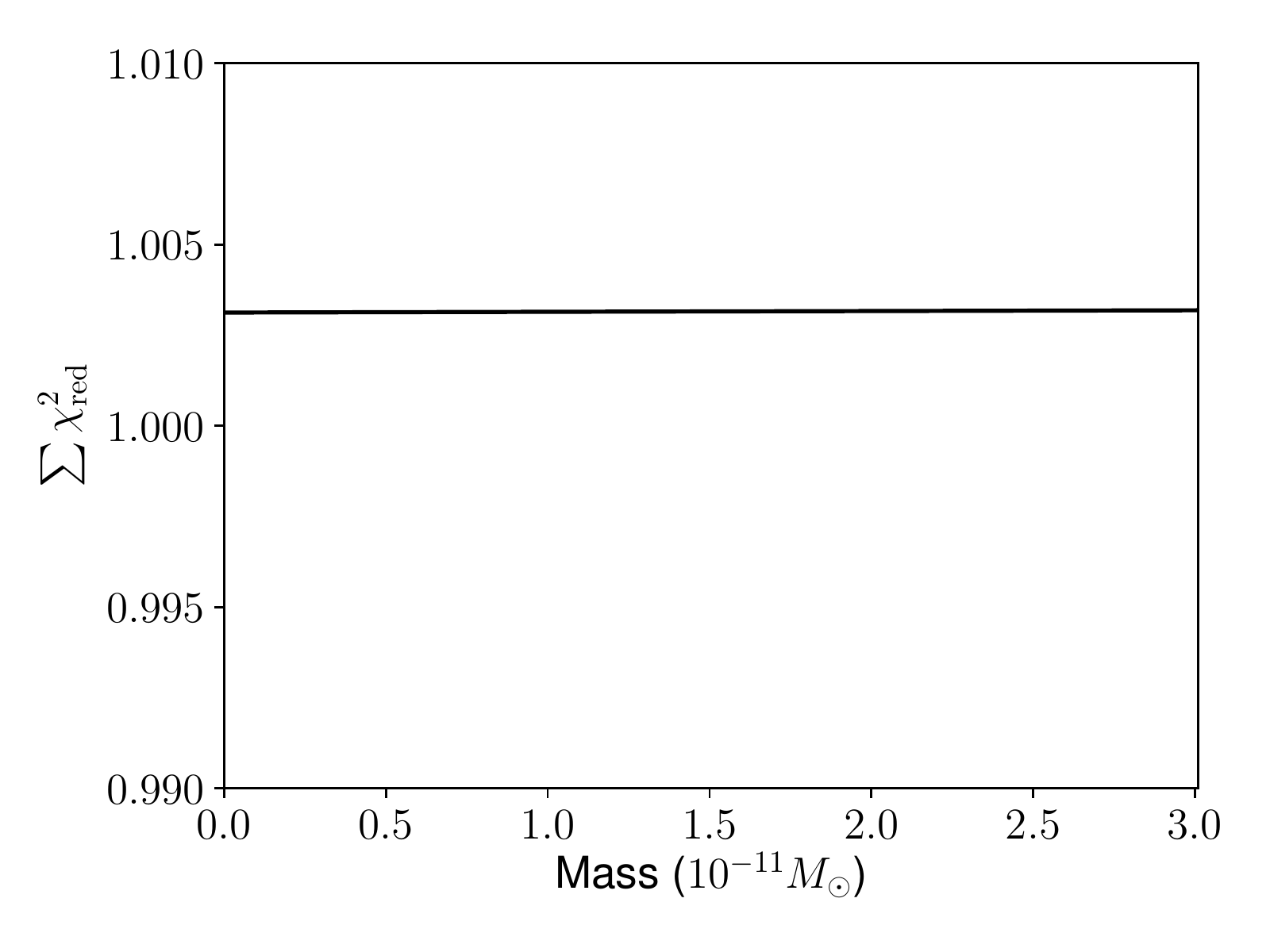}
    \caption{Results of the mass marching algorithm applied to
      [14328;4665].}
    \label{granvik-muinonen}
  \end{center}
\end{figure}

The case of [15;14401], however, is an example of {a
  problematic case for} the marching algorithm in that the algorithm
finds a best-fit mass of zero {(Fig.~\ref{15_march})}.
To verify that the method performs correctly in a case where zero mass
is expected, we ran the marching algorithm for an arbitrary pair of
asteroids, [14328;4665], that do not experience a close
encounter. This being the case, it logically follows that perturber
mass should have no effect on the $\chi^2$ value. This is exactly the
result we see in Fig.~\ref{granvik-muinonen}.
{Taken together, the above results imply that a correlation
  between the mass of the perturber and $\chi^2$ indicates that the
  close encounters has potential for useful mass estimation, even when
  the best-fit mass is zero.}

The best-fit masses for all cases can be seen in Table
\ref{march_compilation}. The best-fit mass is zero in four cases {
  (all show a correlation between mass and $\chi^2$)}, while the other
five give results fairly similar to literature values.
\begin{table}
  \caption{Compilation of the marching algorithm's results for all
    used encounters. The reference masses are from \citet{Car12}.}
  \label{march_compilation}
  \begin{center}
    \begin{tabular}{ccc}
      \hline
      Encounter  &  Marching result      & Ref.~mass \\
      &  [$10^{-11} \msun$] & [$10^{-11} \msun$] \\
      \hline
      {[7;17186]}  &  $0.120$						   & $0.649 \pm 0.106$									   	 \\ 
      {[10;3946]}  &  $2.02$ 							 & $4.34  \pm 0.26$ 											   \\ 
      {[13;14689]} &  $0.00$       	  		 & $0.444 \pm 0.214$									     \\ 
      {[15;14401]} &  $0.00$        			 & $1.58 \pm 0.09$ 					   \\ 
      {[19;3486]}  &  $0.307$ 						 & $0.433 \pm 0.073$   				 \\ 
      {[19;27799]} &  $0.307$ 						 & $0.433 \pm 0.073$    \\ 
      {[29;987]}   &  $0.00$      				 & $0.649 \pm 0.101$    \\ 
      {[52;124]}   &  $1.45$ 	 					   & $1.20 \pm 0.29$    \\ 
      {[704;7461]} &  $0.00$          		 & $1.65 \pm 0.23$ \\ 
      \hline
    \end{tabular}
  \end{center}
\end{table}

\begin{figure}
  \begin{center}
    \includegraphics[width=1.0\columnwidth]{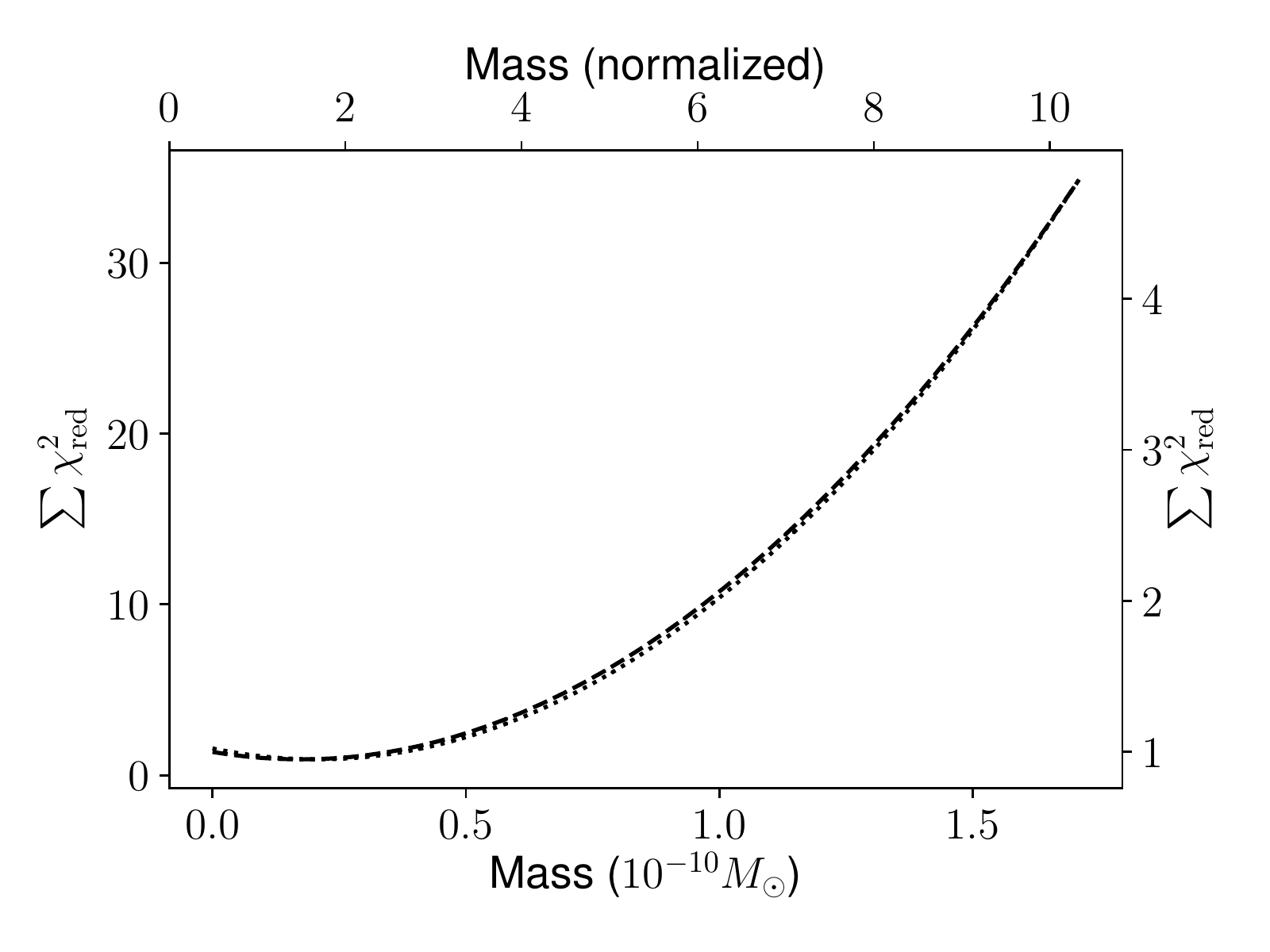}
    \caption{{Results of the mass marching algorithm applied to
        the [15;14401] encounter using two separate initial orbits
        calculated with pre- and post-encounter astrometry
        respectively. The dashed line and the left y-axis correspond
        to the pre-encounter orbit while the dotted line and the right
        y-axis correspond to the post-encounter orbit.}}
    \label{15_both}
  \end{center}
\end{figure}

Given the nature of the marching algorithm, it is logical that the
zero mass for [15;14401] resulted from inaccurate initial orbits,
which did not take the perturbating asteroid's non-zero mass into
account. {To verify this explanation, we calculated two
  separate initial orbits.  The first used pre-encounter astrometry
  only and the second used post-encounter astrometry only.  We then
  ran the marching algorithm separately for both orbits while using
  all of our available astrometry. Both of these initial orbits result
  in a non-zero minimum (Fig.~\ref{15_both}) unlike the initial orbit
  that used all of our astrometry. This confirms that the zero minimum
  was indeed caused by inaccurate initial orbits. In addition, the
  curves have very similar shapes, although the $\chi^2_\mathrm{red}$
  values are quite different.}

\subsection{Results for the Nelder-Mead algorithm}

The Nelder-Mead algorithm fits for both the best mass and orbital
elements.  We first ran the Nelder-Mead algorithm for the synthetic
data.  The best fit mass found was $6.94 \times 10^{-11}\msun$, which
is slightly lower than the marching result and, in this case,
approximately 78\% of the correct value, which is a reasonable
result. As one would expect, the fit is significantly better than that
of the marching algorithm, as the marching algorithm's best
{$\chi^2_\mathrm{red}$ value is 2.17, while the best
  Nelder-Mead fit had a $\chi^2_\mathrm{red}$ value of 1.16.}

\begin{figure}
  \begin{center}
    \includegraphics[width=1.0\columnwidth]{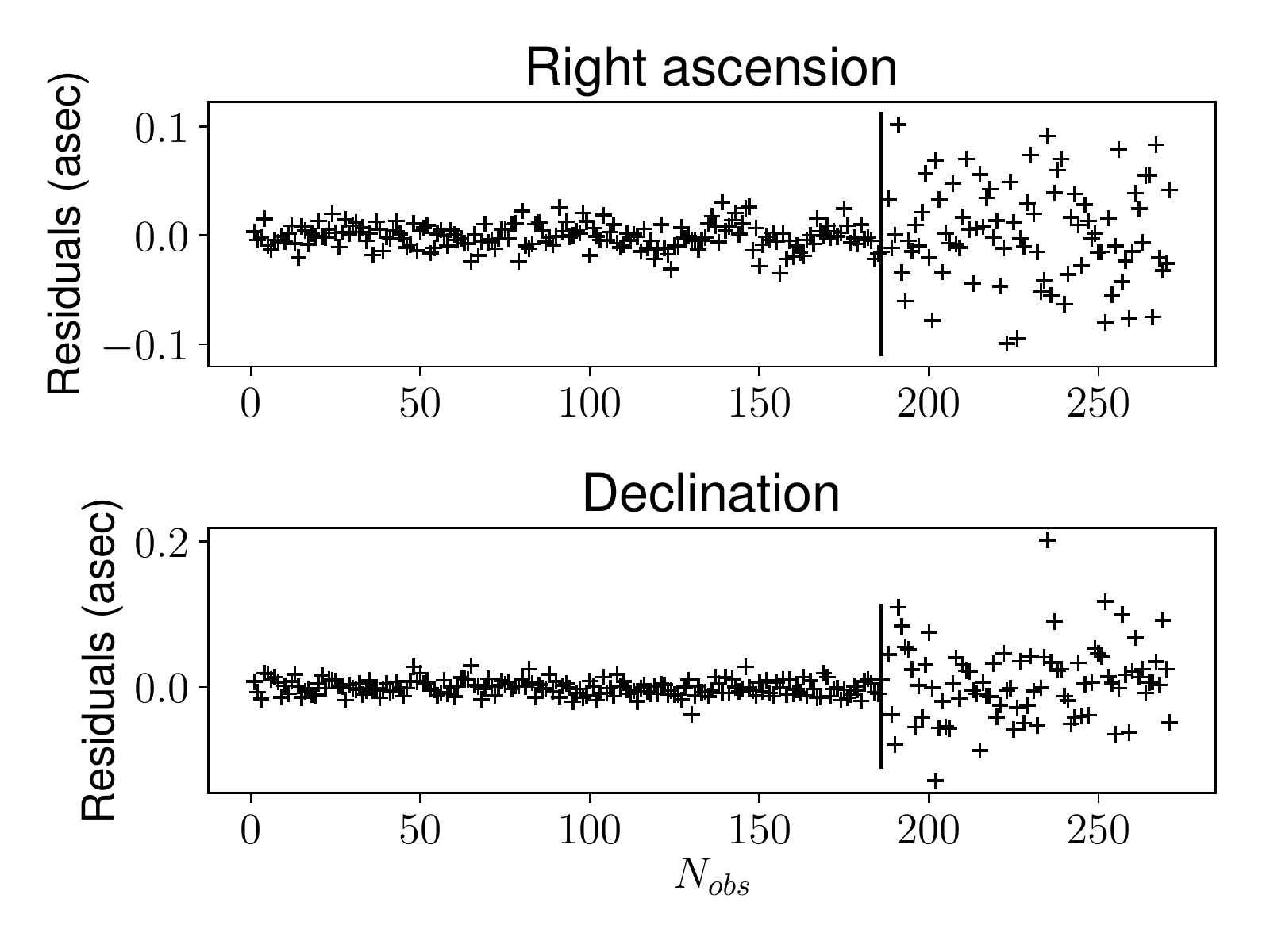}
    \caption{Residuals of the best fit for synthetic data obtained
      with the Nelder-Mead method. The data for the perturber and the
      test asteroid are separated by the vertical line, with the
      perturber's data on the left side. The data points are sorted by
      observation time in such a manner that each asteroid's data
      begins with the earliest observation and ends with the last.}
    \label{synth_simplex_residuals}
  \end{center}
\end{figure}

Residuals for the Nelder-Mead method are shown in
Fig.~\ref{synth_simplex_residuals}. The residuals still appear
reasonable, and in comparison to the marching residuals shown in
Fig.~\ref{synth_march_residuals}, the Nelder-Mead residuals are
clearly lower and the systematic trend in the right ascension is gone
due to the algorithm improving the used orbits themselves. For
comparison, the test asteroid's RMS values from the best fit of the
marching algorithm were 0.058" and 0.061" for the right ascension and
declination respectively while the equivalent values from the
Nelder-Mead run were 0.044" and 0.052", which are clearly lower values
and thus a better fit.
Results for the real encounters can be seen in
Table~\ref{neldermead}. In most cases, the result is again similar to
literature values. Notable exceptions include [13;14689], [29;987], in
which the result is too large, and [52;124], where it is too small. Of
course, one has to keep in mind that these estimates do not include
uncertainties.
\begin{table}
  \caption{Compilation of the Nelder-Mead algorithm's results for the
    real encounters considered. The reference masses are from
    \citet{Car12}.}
  \label{neldermead}
  \begin{center}
    \begin{tabular}{ccc}
      \hline
      Encounter  & Nelder-Mead result                & Ref.~mass \\
      & [$10^{-11} \msun$] & [$10^{-11} \msun$] \\
      \hline
	  {[7;17186]}  & 0.503 								       & $0.649 \pm 0.106$   									  \\ 
	  {[10;3946]}  & 2.76  										   & $4.34 \pm 0.26$											    \\ 
	  {[13;14689]} & 1.37  											 & $0.444 \pm 0.214$										    \\ 
	  {[15;14401]} & 3.29  					  					 & $1.58 \pm 0.09$											    \\ 
	  {[19;3486]}  & 0.232  										 & $0.433 \pm 0.073$ 									   \\ 
	  {[19;27799]} & 0.220 											 & $0.433 \pm 0.073$ 									   \\ 
	  {[29;987]}   & 1.43												 & $0.649 \pm 0.101$ 									   \\ 
	  {[52;124]}   & 0.774  								     & $1.20 \pm 0.29$ 										   \\ 
	  {[704;7461]} & 2.98 											 & $1.65 \pm 0.23$ \\ 
          \hline
    \end{tabular}
  \end{center}
\end{table}

\subsection{MCMC algorithm results}

\begin{figure}
  \begin{center}
    \includegraphics[width=1.0\columnwidth]{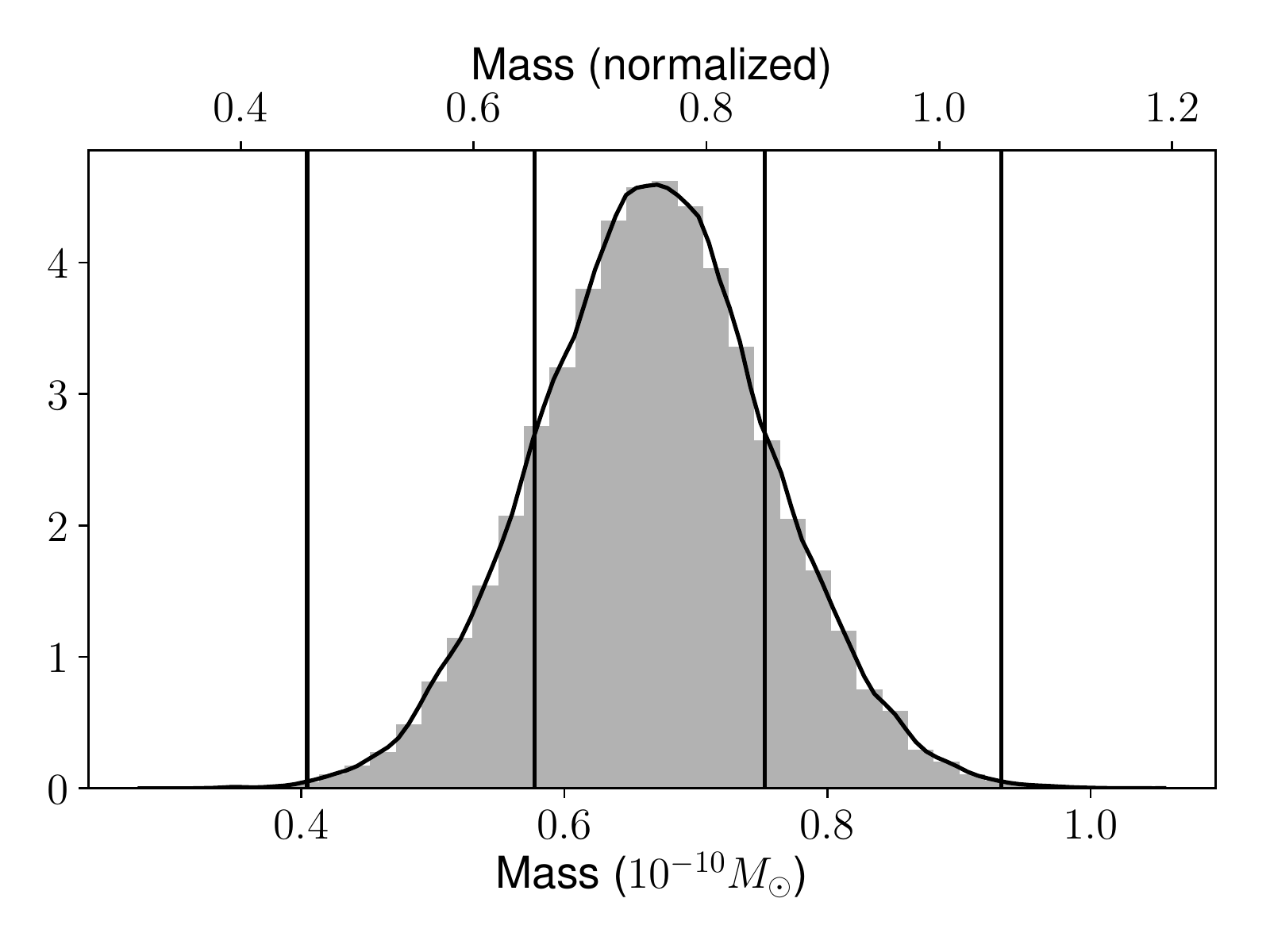}
    \caption{A histogram of masses for all transitions of the MCMC
      chain for synthetic data. Each mass is weighted by the total
      amount of repetitions for that particular transition, i.e., if a
      set of parameters is accepted $n$ times in a row, it will be
      counted $n$ times into the histogram. The bottom $x$ axis
      represents mass in solar mass $\msun$, while the top $x$ axis is
      normalized such that 1.0 represents the correct mass. The black
      graph represents a kernel density estimate while the
      {vertical} lines represent the $1\sigma$ and $3\sigma$
      confidence limits. The $y$ axis is normalized such that the
      kernel density estimate's integral over the whole $x$ axis is
      one.}
    \label{synth_mcmc_mass}
  \end{center}
\end{figure}

We ran the MCMC algorithm for both the synthetic case and all of the
real encounters considered for a total of 50,000 transitions.
Fig.~\ref{synth_mcmc_mass} displays the resulting probability
distribution of perturber mass for the synthetic test case. Upon
examination of the kernel density estimate, which estimates the
probability density function of the mass, one can see that the best
fitting mass is approximately $6.71 \times 10^{-11}\msun$, or 76\% of
the correct mass of $8.85 \times 10^{-11}\msun$. Interestingly, the
result is very close to our Nelder-Mead results.  The correct result
is within our $3\sigma$ confidence limits. It is also apparent that
the probability distribution in this case is essentially Gaussian.

\begin{figure}
  \begin{center}
    \includegraphics[width=1.0\columnwidth]{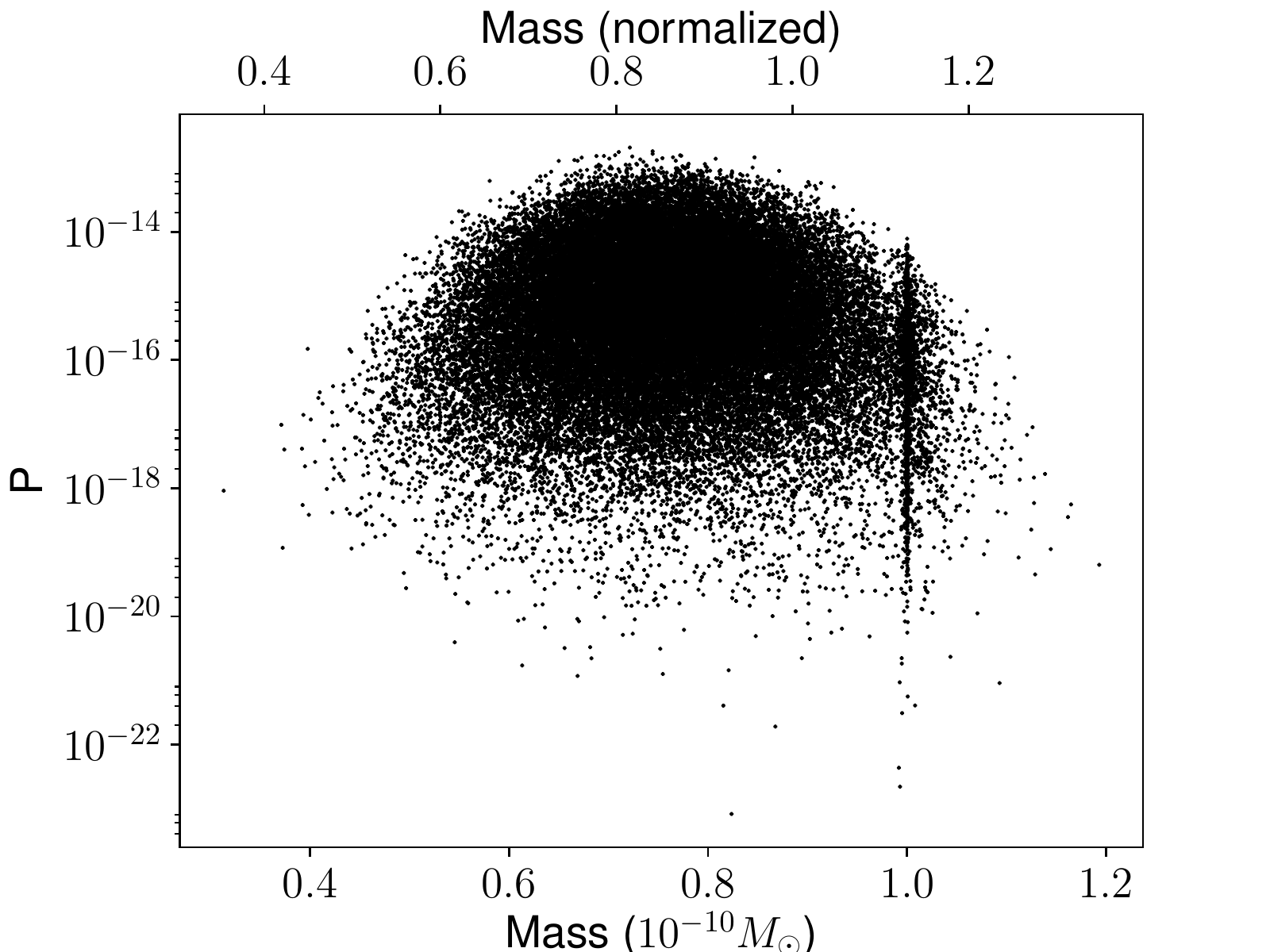}
    \caption{Scatter plot of mass versus probability density value for
      all transitions of the MCMC chain for synthetic data. The
      vertical clump of data points at roughly $10^{-10}\msun$
      corresponds to the initial burn-in phase where the proposal
      distribution is far too narrow.}
    \label{synth_masspdv}
  \end{center}
\end{figure}

For comparison, we have also included a logarithmic scatter plot of
mass versus the probability density value for each transition
(Fig.~\ref{synth_masspdv}). It is apparent that the scatter plot
matches the distribution quite well, and one can also see that the
distributions for both halves of the chain are quite similar. These
result are to be expected, and shows that there appear to be no major
issues with the chain itself.

\begin{figure}
  \begin{center}
    \includegraphics[width=1.0\columnwidth]{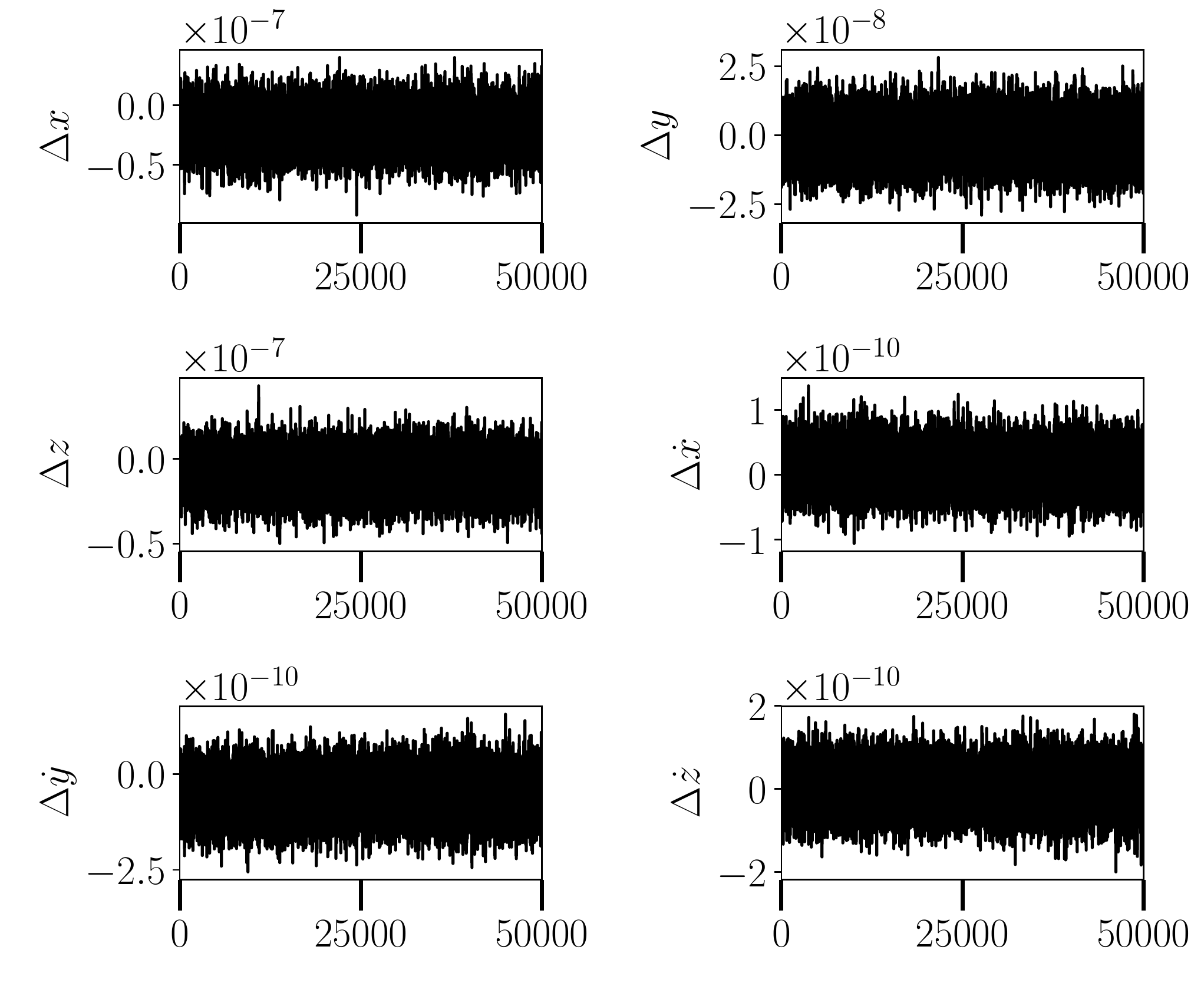}
    \caption{The trace of the MCMC chain in terms of perturber orbital
      elements when using the synthetic astrometry. The $x$-axis
      displays the number of the transition, while the $y$-axis
      represents difference from the initial orbit.{The units
        are in au and au/day.}}
    \label{synth_perturber_elems_evo}
  \end{center}
\end{figure}

\begin{figure}
  \begin{center}
    \includegraphics[width=1.0\columnwidth]{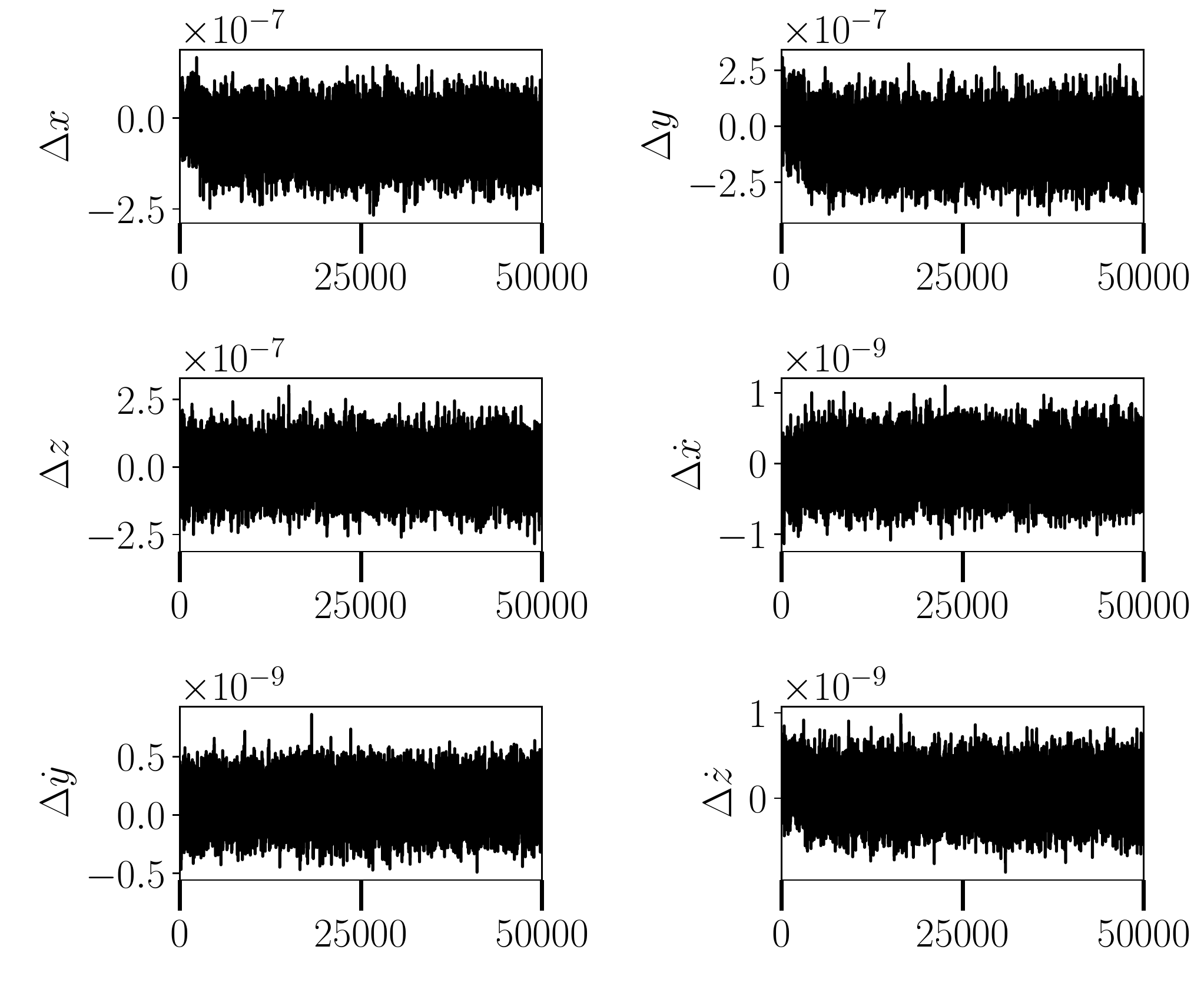}
    \caption{The trace of the MCMC chain in terms of test asteroid
      orbital elements in the synthetic case. The $x$-axis displays
      the number of the transition, while the $y$-axis represents
      difference from the initial orbit. {The units are in au
        and au/day.}}
    \label{synth_test_elems_evo}
  \end{center}
\end{figure}

Figures~\ref{synth_perturber_elems_evo} and \ref{synth_test_elems_evo}
display the trace of the orbital elements for both the perturbing
asteroid and the test asteroid. The algorithm converges very fast,
with no visible burn-in period on this scale, and there are no mixing
issues to be seen. Overall the figures look exactly like what one
would expect from a good MCMC chain.

\begin{figure}
  \begin{center}
    \includegraphics[width=1.0\columnwidth]{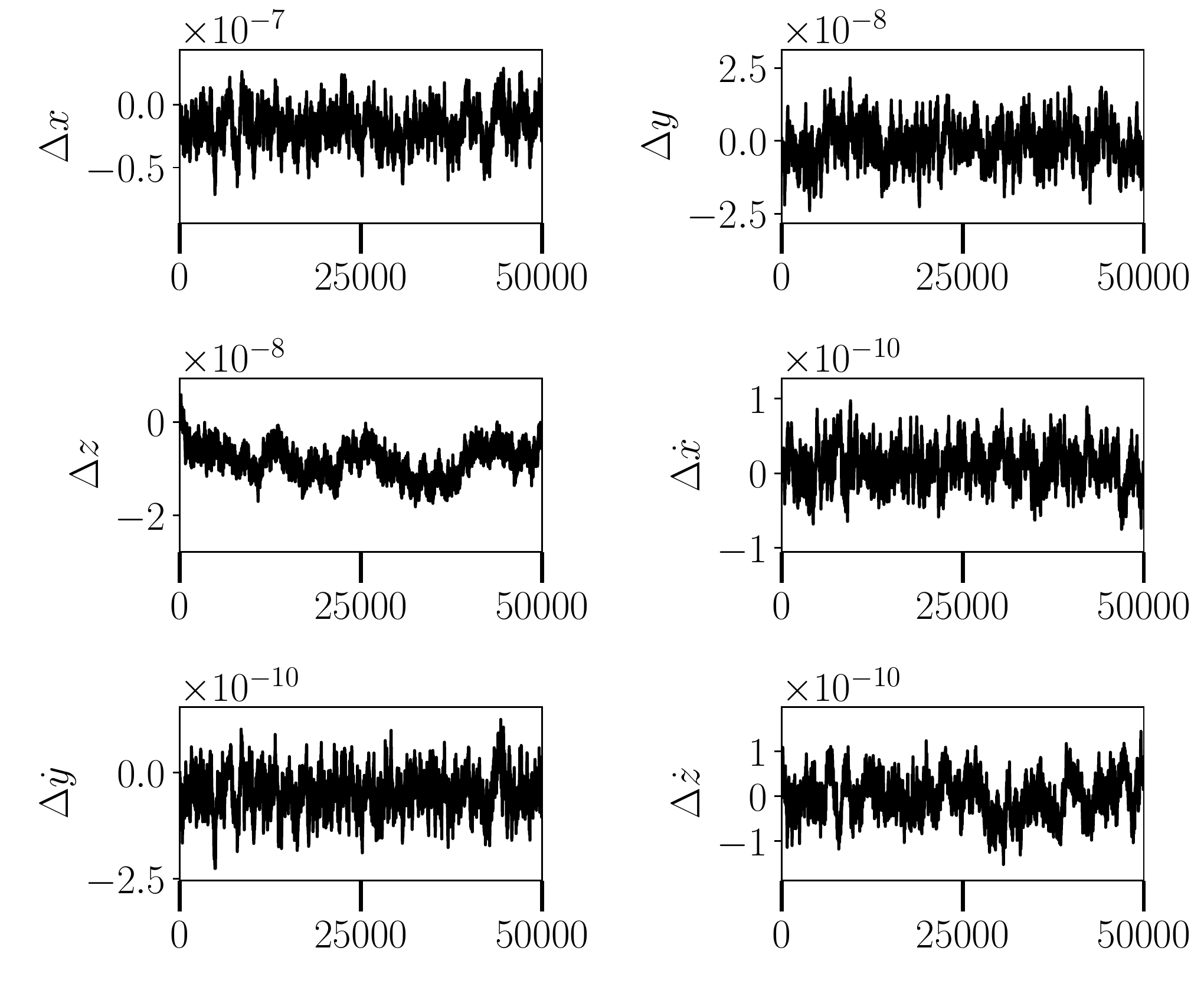}
    \caption{The trace of the MCMC chain in terms of perturber orbital
      elements for a run without Adaptive Metropolis using the
      synthetic astrometry. There are visible convergence and mixing
      issues due to poor initial proposal distributions. The $x$-axis
      displays the number of the transition, while the $y$-axis
      represents difference from the initial orbit. {The units
        are in au and au/day.}}
    \label{synth_perturber_noam}
  \end{center}
\end{figure}

\begin{figure}
  \begin{center}
    \includegraphics[width=1.0\columnwidth]{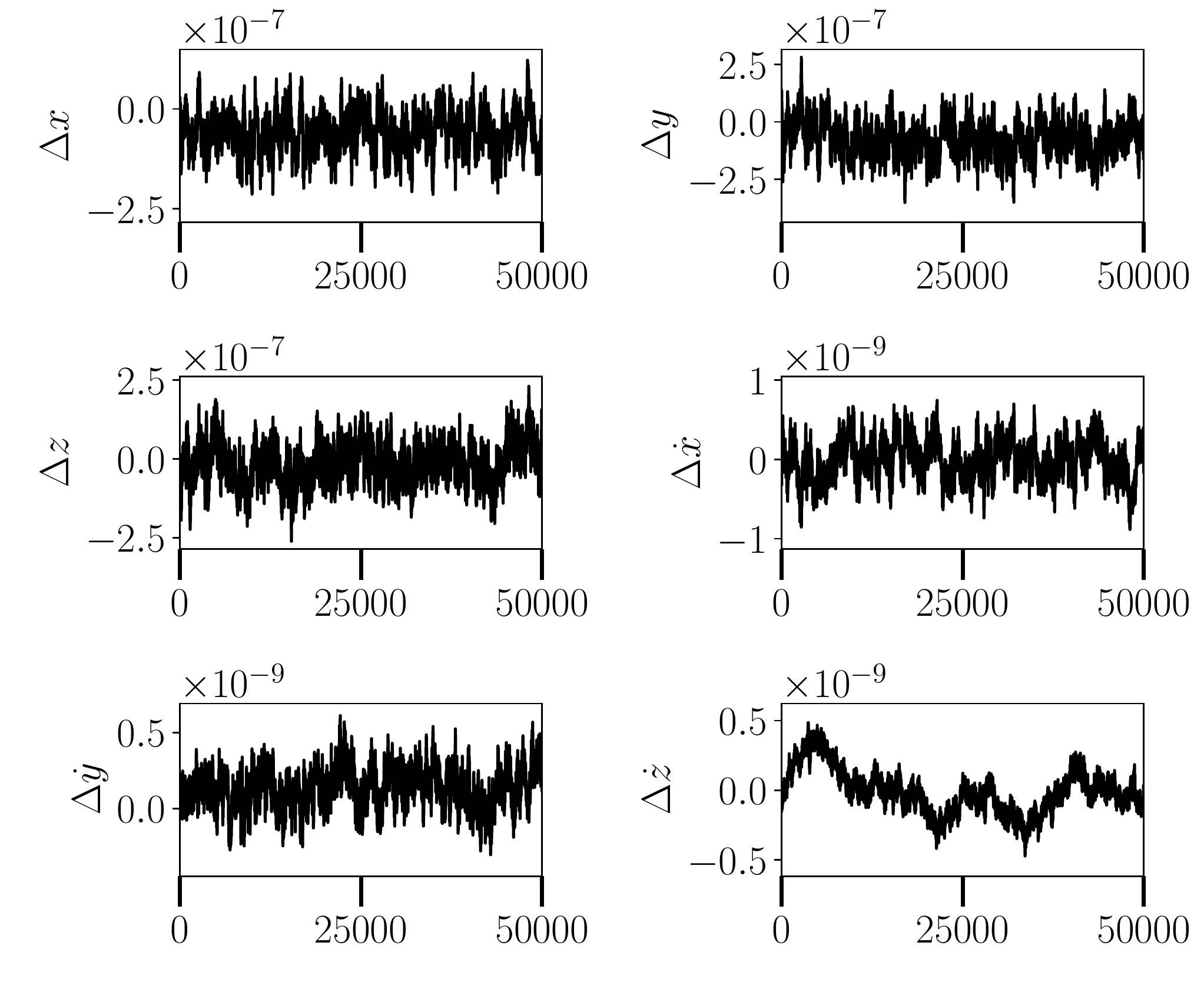}
    \caption{The trace of the MCMC chain in terms of test asteroid
      orbital elements for a run without Adaptive Metropolis using the
      synthetic astrometry. There are visible convergence and mixing
      issues due to poor initial proposal distributions. The $x$-axis
      displays the number of the transition, while the $y$-axis
      represents difference from the initial orbit. {The units
        are in au and au/day.}}
    \label{synth_test_noam}
  \end{center}
\end{figure}

To showcase the usefulness of AM, we have included equivalent plots
showing the evolution of each orbital element in the synthetic test
case without AM in Figs~\ref{synth_perturber_noam} and
\ref{synth_test_noam}. Upon visual examination it is immediately
apparent that mixing is poor in all cases, and $z$ of the perturber
and $\dot{z}$ of the test asteroid do not converge properly.  When
compared with the equivalent Figs.~\ref{synth_perturber_elems_evo} and
\ref{synth_test_elems_evo} where AM was used, the difference is
substantial. It is clear that AM by itself has completely negated
these problems and thus greatly improved our algorithm.

\begin{figure}
  \begin{center}
    \includegraphics[width=1.0\columnwidth]{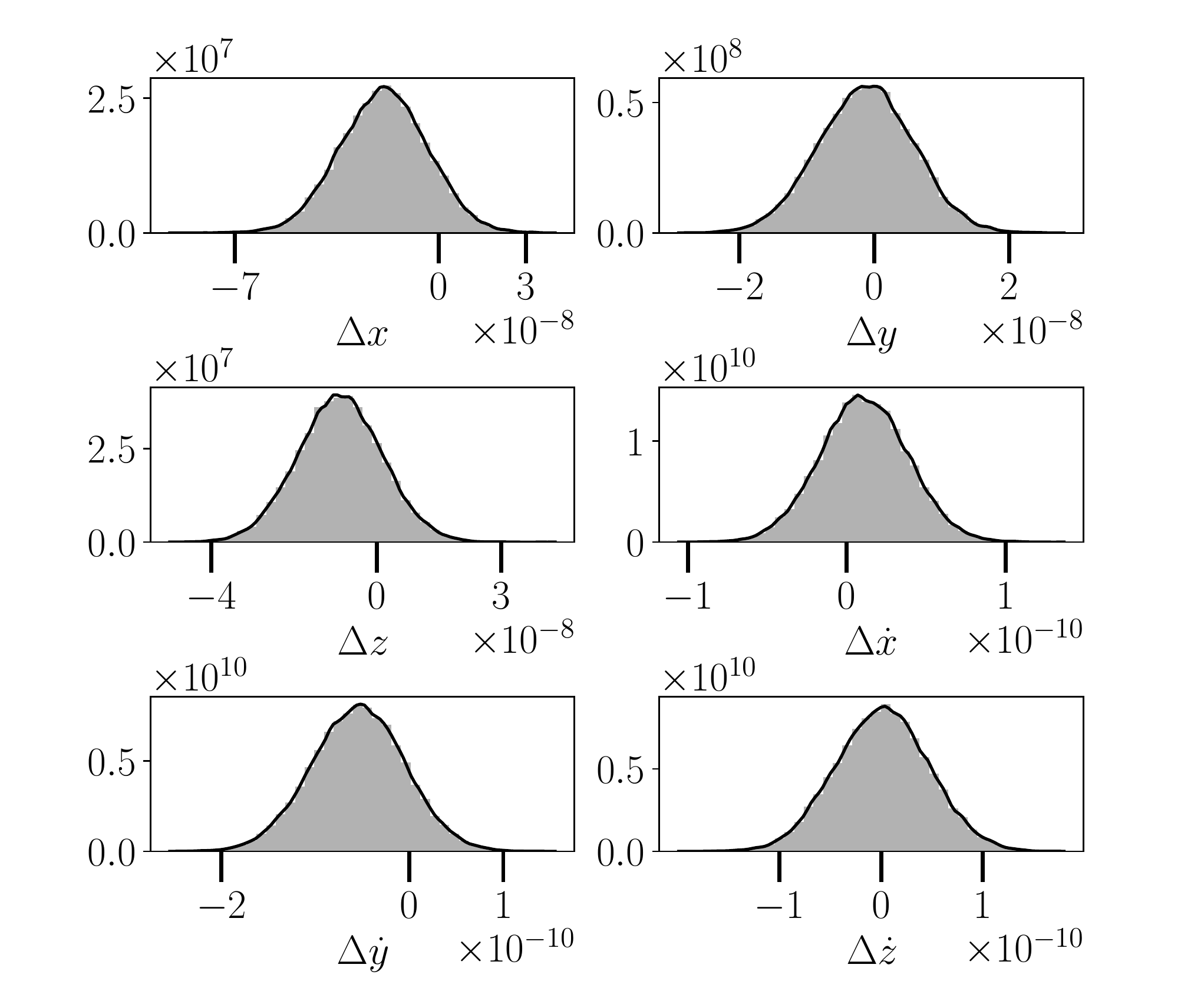}
    \caption{Histograms of the Cartesian state vectors for the
      perturber for all transitions of the MCMC chain for synthetic
      data. Each element is weighted by the total amount of
      repetitions for that particular transition, i.e. if a proposal
      is accepted $n$ times in a row, it will be counted $n$ times
      into the histogram.  The bottom $x$ axis represents difference
      between the initial synthetic orbit and the tested orbit, which
      means that 0.0 corresponds to the initial orbit.  {The
        units are in au and au/day.}  The black graph represents a
      kernel density estimate while the $y$ axis is normalized such
      that the kernel density estimate's integral over the whole $x$
      axis is one.}
    \label{synth_perturber_elems}
  \end{center}
\end{figure}

\begin{figure}
  \begin{center}
    \includegraphics[width=1.0\columnwidth]{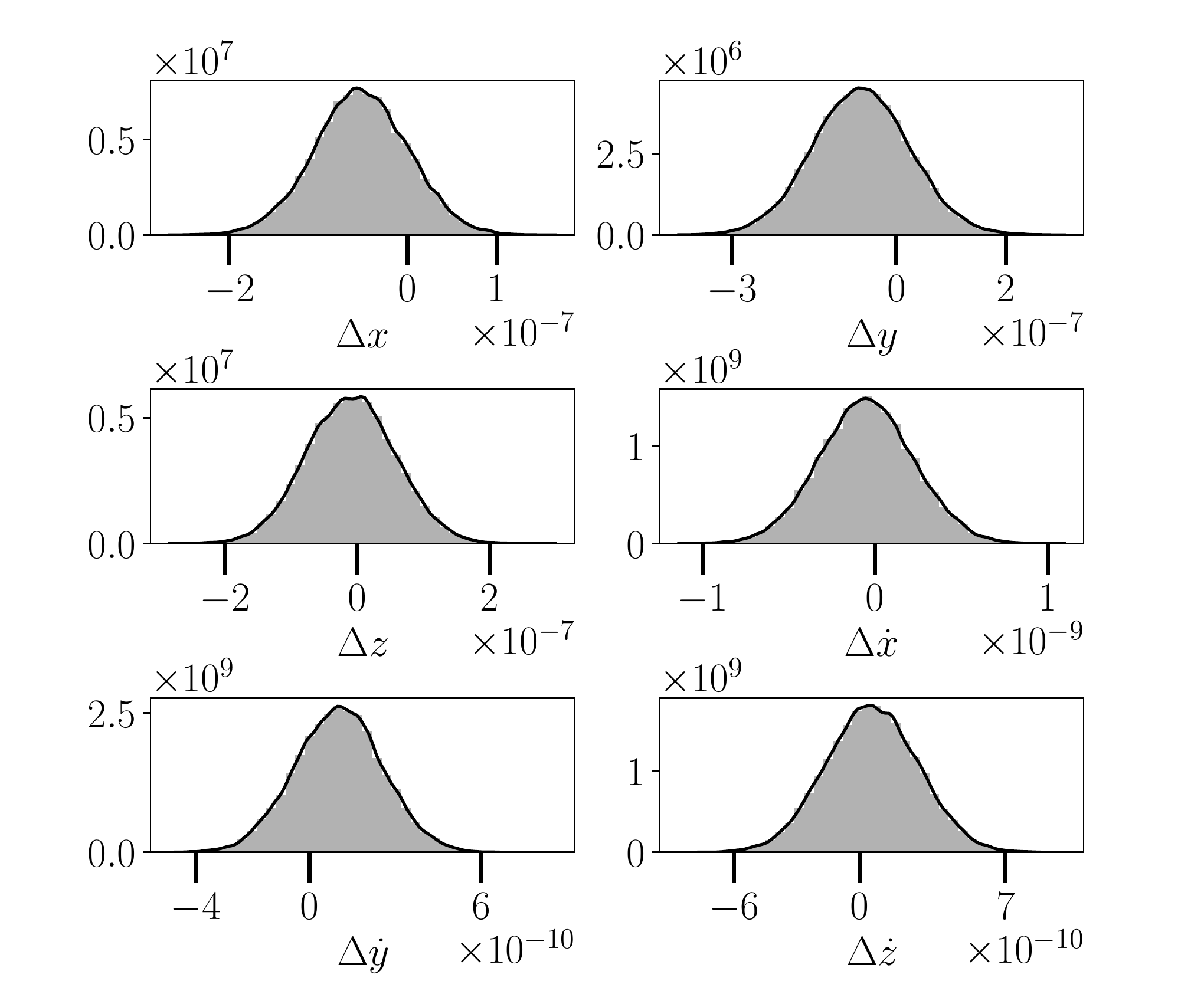}
    \caption{Histograms of Cartesian state vectors for the test
      asteroid for all transitions of the MCMC chain for synthetic
      data. Each element is weighted by the total amount of
      repetitions for that particular proposal, i.e. if a proposal is
      accepted $n$ times in a row, it will be counted $n$ times into
      the histogram.  The bottom $x$ axis represents difference
      between the initial synthetic orbit and the tested orbit, which
      means that 0.0 corresponds to the initial orbit.  {The
        units are in au and au/day.}  The black graph represents a
      kernel density estimate while the $y$ axis is normalized such
      that the kernel density estimate's integral over the whole $x$
      axis is one.}
    \label{synth_test_elems}
  \end{center}
\end{figure}

Finally, Figs.~\ref{synth_perturber_elems} and \ref{synth_test_elems}
show the distributions of the orbital elements for both the perturber
and test asteroid in the case of synthetic astrometry. The
distributions also appear to be largely Gaussian and quite narrow.

\begin{figure}
  \begin{center}
    \includegraphics[width=1.0\columnwidth]{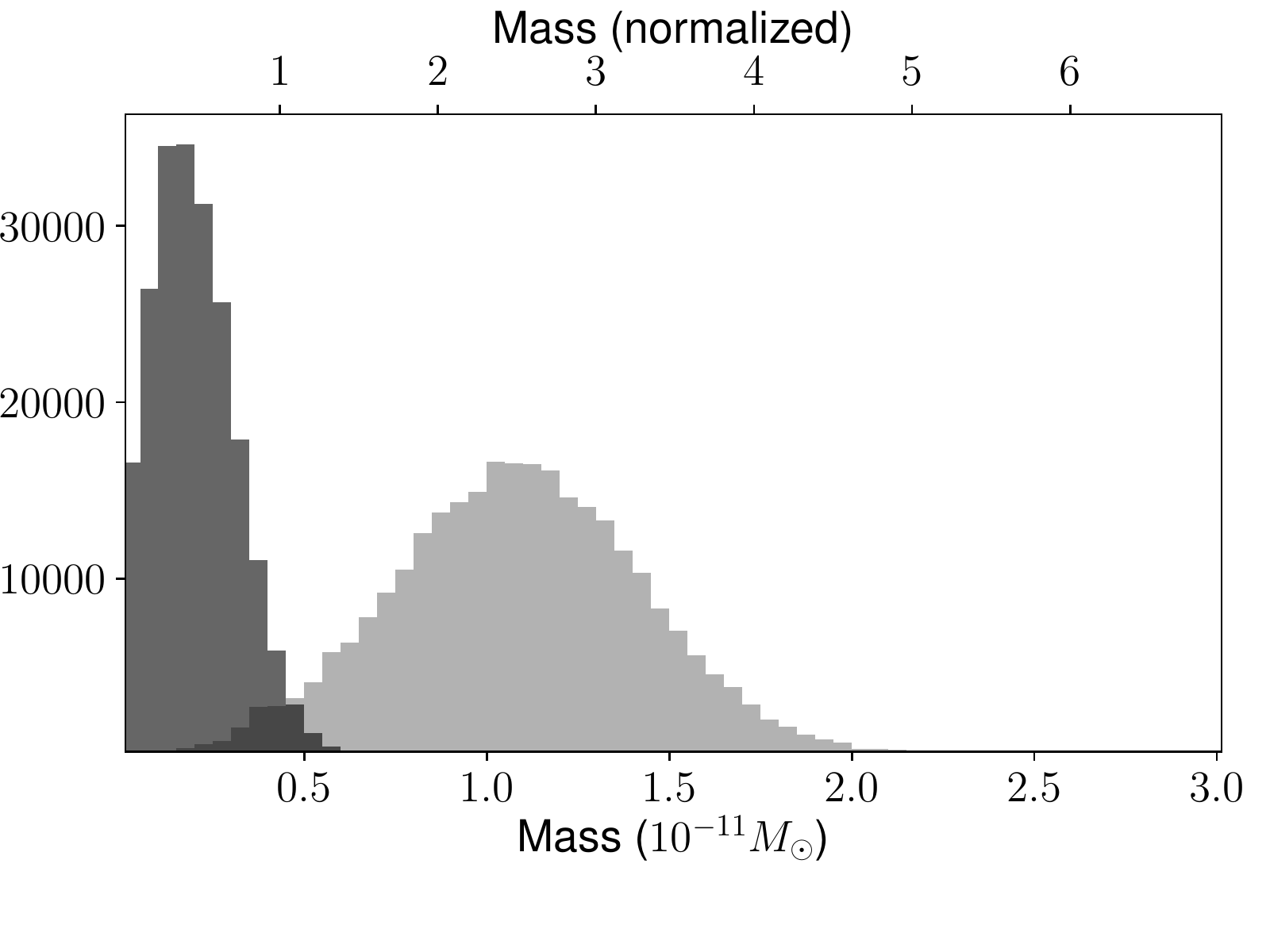}
    \caption{Results of the MCMC algorithm applied to the [19;3486]
      (darker shade) and [19;27799] encounters. The upper x-axis is
      normalized such that 1.0 equals the literature value of
      \citet{Car12}.}
    \label{19_masses}
  \end{center}
\end{figure}

Example results for real asteroids are shown in
Fig.~\ref{19_masses}. The figure shows the resulting mass probability
distributions for the [19;3486] and [19;27799] encounters. Note that
in both cases, the perturbing asteroid is the same, and thus the
correct mass should also be the same. However, the choice of the test
asteroid has a significant impact on the results due to observational
errors and systematic biases. Such a result is not unprecedented: for
instance \citet{Bae11} obtained a value of $(3.90 \pm 0.37) \times
10^{-12}\msun$ for [19;3486] and $(9.10 \pm 1.6) \times 10^{-12}\msun$
for [19;27799], a roughly 2x difference. The value of \citet{Car12} is
a weighted average of all mass estimates for this asteroid and as
such, is not dependent on only a single test asteroid.
One can see that our maximum-likelihood solutions for either pair do
not perfectly correspond to the literature value of \citet{Car12}, but
it remains well within the 3-sigma limits of both. Intriguingly, the
literature value for the mass corresponds well to the peak of the area
where both histograms overlap. This suggests that if we were to use
both test asteroids simultaneously, we would get results much closer
to the literature value. This is encouraging, as we intend to extend
our algorithm to use multiple test asteroids and/or perturbers
simultaneously.  The Nelder-Mead algorithm found masses of $2.32
\times 10^{-12}\msun$ and $2.20 \times 10^{-12}\msun$ for these
encounters respectively, which intriguingly are much closer to each
other than the MCMC results.

\begin{figure}
  \begin{center}
    \includegraphics[width=1.0\columnwidth]{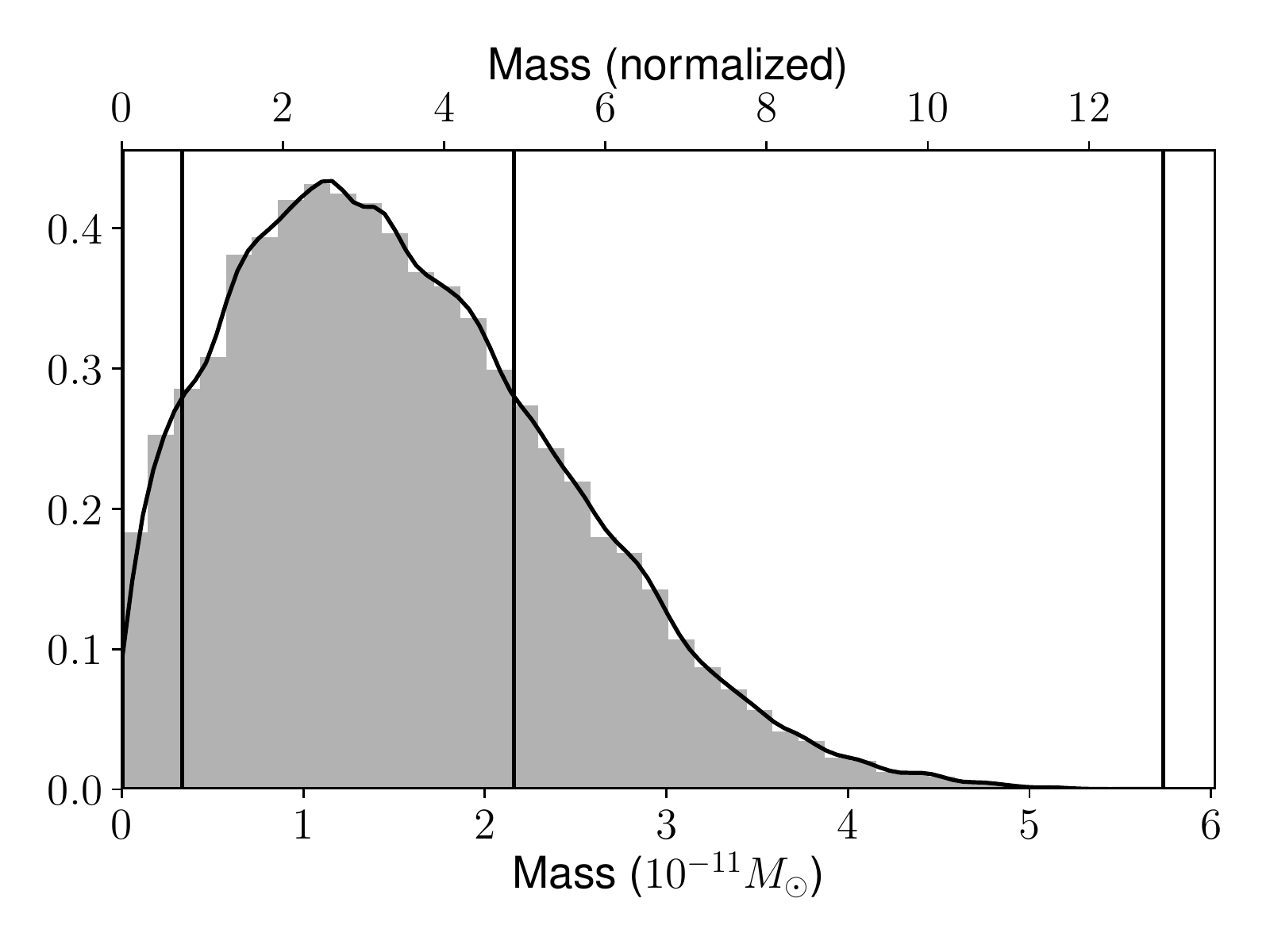}
    \caption{Results of the MCMC algorithm applied to the [13;14689]
      encounter. The upper x-axis is normalized such that 1.0 equals
      the literature value of \citet{Car12}.}
    \label{13_mass}
  \end{center}
\end{figure}

Figure~\ref{13_mass} shows a case where the probability distribution
of perturber mass is clearly non-Gaussian, thus showing that Gaussian
error estimates are incorrect for this case. Again our
maximum-likelihood mass does not correspond to the literature value,
but is still well within uncertainty limits.

\begin{figure}
  \begin{center}
    \includegraphics[width=1.0\columnwidth]{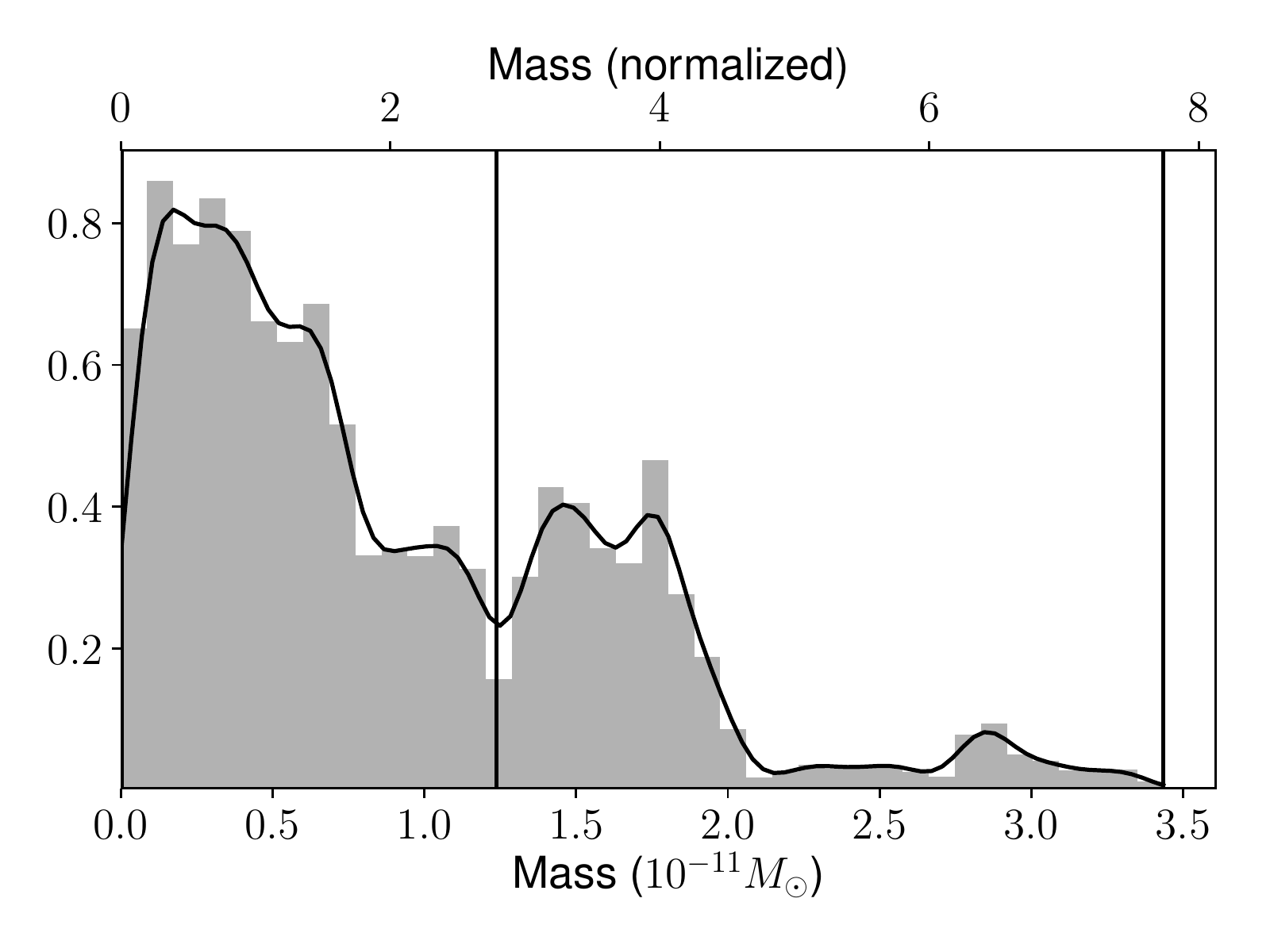}
    \caption{Results of the MCMC algorithm without Adaptive Metropolis
      applied to the [13;14689] encounter. The upper x-axis is
      normalized such that 1.0 equals the literature value of
      \citet{Car12}.}
    \label{13_noam}
  \end{center}
\end{figure}

[13;14689] is also the only case where AM had a significant impact on
the mass probability distribution. For comparison, the equivalent plot
of a run of the same length without AM is shown in
Fig.~\ref{13_noam}. Clearly, in this case mass does not converge
properly, likely due to poor convergence and mixing of the orbits. In
our other cases, the impact of AM on perturber mass was much smaller
despite the significant improvement on mixing and convergence of the
orbital elements.

\begin{table*} 
  \caption{Compilation of the MCMC algorithm's results for all used
    encounters. The reference masses are from \citet{Car12}.}
  \label{mcmc_compilation}
  \begin{center}
    \begin{tabular}{ccccc}
      \hline
      Encounter	  & ML mass              & $1\sigma$ boundaries   & $3\sigma$ boundaries   & Ref.~mass \\
      & [$10^{-11} \msun$] & [$10^{-11} \msun$]   & [$10^{-11} \msun$]   & [$10^{-11} \msun$]  \\
      \hline       
      Synthetic   & 6.71 & [5.84, 7.62]  & [4.09, 9.43] & 8.85 $\pm$ 0.00 \\
      {[7;17186]} & 0.210  & [0.0499, 0.354] & [0.000660, 1.20]  & 0.649 $\pm$ 0.106 \\
      {[10;3946]} & 2.48 & [2.21, 2.77] & [1.63, 3.32]  & 4.34 $\pm$ 0.26 \\
      {[13;14689]}	& 1.13	 & [0.332, 2.16]  & [0.00273, 5.74] & 0.444 $\pm$ 0.214 \\
      {[15;14401]} 	& 1.11	 & [0.914, 1.25]  & [0.574, 1.61] & 1.58 $\pm$ 0.09 \\
      {[19;3486]} 	& 0.141 & [0.0567, 0.285]  & [0.000467, 0.828]      & 0.433 $\pm$ 0.073	\\
      {[19;27799]} 	& 1.11	& [0.757, 1.42]  & [0.122, 2.05] & 0.44 $\pm$ 0.073	\\
      {[29;987]}	& 0.258	& [0.0163, 0.898]  & [0.00238, 4.43] & 0.649 $\pm$ 0.101 \\
      {[52;124]} & 0.893 & [0.232, 1.91]        & [0.00319, 6.05]  & 1.20 $\pm$ 0.29 \\
      {[704;7461]} 	& 0.155 & [0.00910, 0.664]  & [0.00182, 3.50]  & 1.65 $\pm$ 0.23 \\
      \hline
    \end{tabular}
  \end{center}
\end{table*}

Table \ref{mcmc_compilation} shows our MCMC results for all of our
selected encounters along with their uncertainty limits. One can see
that in almost all cases, our maximum-likelihood results do not
perfectly correspond to the weighted average values of \citet{Car12}.
Nonetheless, in almost all cases the literature values are well within
our $3\sigma$ confidence limits and within $1\sigma$ in several
cases. These confidence limits are also quite wide in comparison to
those in literature and, significantly, in many cases
non-Gaussian. This can be seen by examining the uncertainty limits:
for Gaussian probability distributions, the $3\sigma$ uncertainty
limits should be three times higher than the equivalent $1\sigma$
limits. As an example, in the case of [704;7461] the upper $3\sigma$
limit is in fact approximately 4 times the $1\sigma$ limit, which
means that the distribution here has a longer tail than a Gaussian
distribution would. On the other hand, the difference between the
lower limits is tiny and also clearly non-Gaussian. The Gaussian cases
of the synthetic pair, [15;14401], and [10;3946] interestingly have
notably smaller confidence limits than the other cases. This suggests
that the symmetric confidence limits typically used in literature are
actually very inaccurate in non-Gaussian cases.

\begin{figure}
  \begin{center}
    \includegraphics[width=1.0\columnwidth]{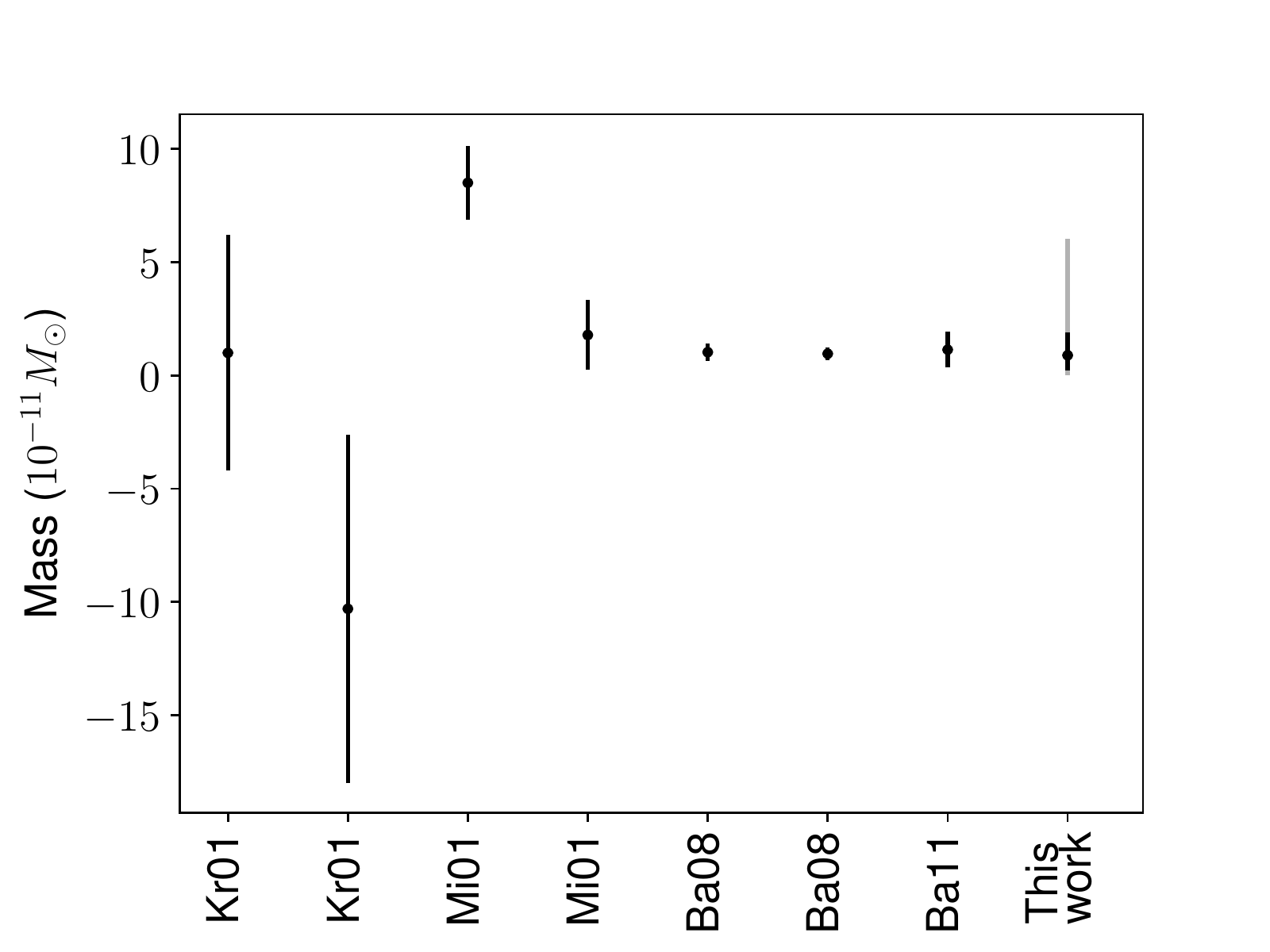}
    \caption{A comparison of our results and previous mass estimates
      {\citep{Kr01,Mi01,Ba08,Bae11}} for (52)~Eunomia done with
      the close encounter method. {Only previous estimates
        computed with a single test asteroid are shown. When results
        for multiple test asteroids had been reported separately, we
        selected the maximum and minimum values.} The rightmost data
      point represents our results, where the black error bars
      represent the $1\sigma$ uncertainty limit while the gray error
      bar represents the $3\sigma$ limit.}
    \label{52_vs_lit}
  \end{center}
\end{figure}

\begin{figure}
  \begin{center}
    \includegraphics[width=1.0\columnwidth]{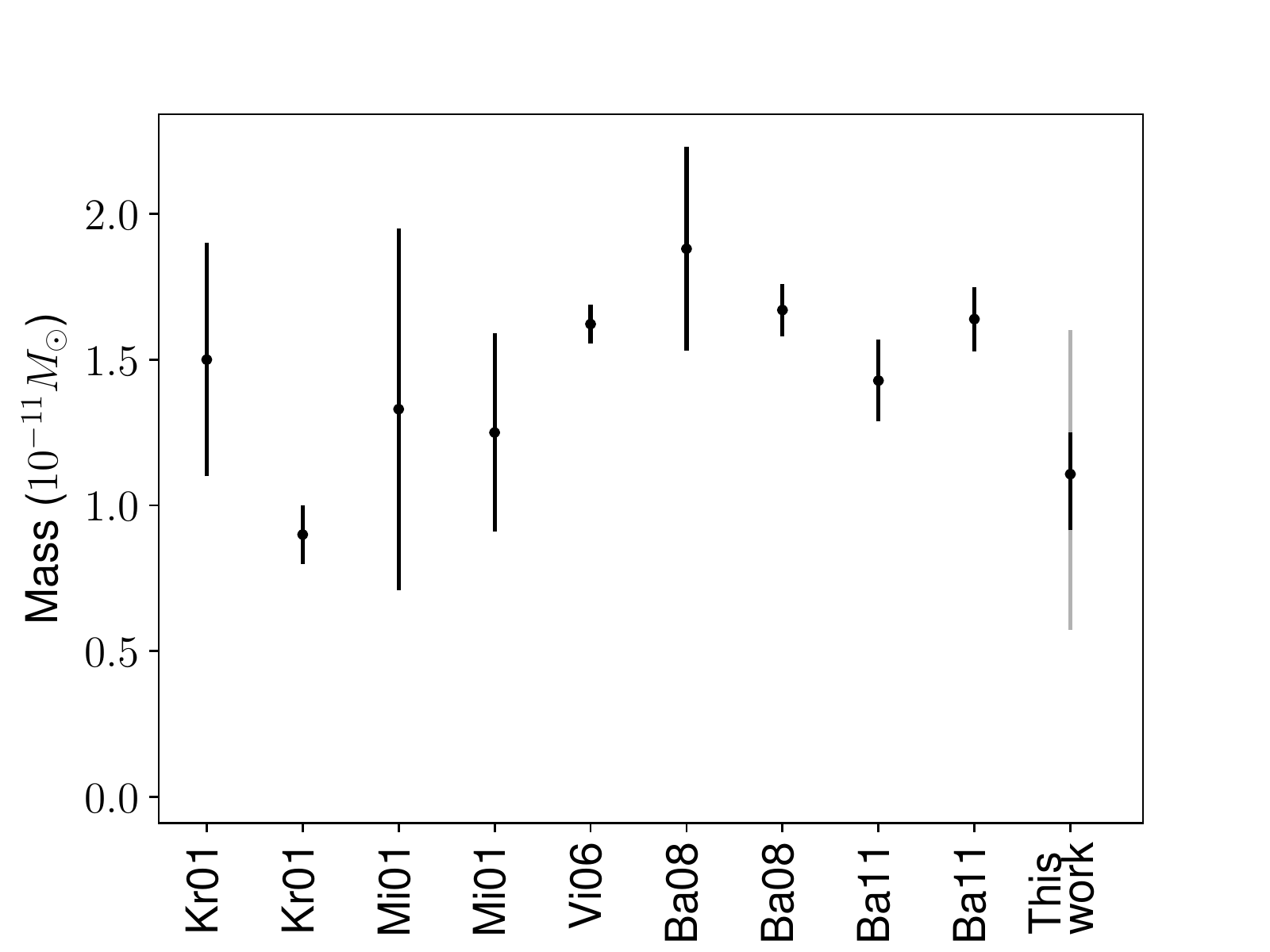}
    \caption{A comparison of our results and previous mass estimates
      {\citep{Kr01,Mi01,Vi06,Ba08,Bae11}} for (15)~Eunomia done
      with the close encounter method. {Only previous estimates
        computed with a single test asteroid are shown. When results
        for multiple test asteroids had been reported separately, we
        selected the maximum and minimum values.} The rightmost data
      point represents our results, where the black error bars
      represent the $1\sigma$ uncertainty limit while the gray error
      bar represents the $3\sigma$ limit.}
    \label{15_vs_lit}
  \end{center}
\end{figure}

Figure \ref{52_vs_lit} shows how our results for (52)~Europa compare
to previous mass estimates for this asteroid.  Interestingly, while
the earlier estimates of this case largely disagree with each other,
{apart from the negative value} every single one of these
estimates falls within our 3$\sigma$ uncertainty limits. This appears
to confirm that previous uncertainties are indeed far too small, and
that our MCMC algorithm provides more realistic uncertainty
estimates. One can also see here that the upper $3\sigma$ limit is
significantly larger than the Gaussian limit would be, which shows
that probability distribution also is not Gaussian in this
case. {The wide uncertainty estimates may also in part be
  explained by the limited amount of pre-encounter astrometry
  used. Nonetheless, our result remains within 1$\sigma$ of the
  \citet{Car12} value.} An equivalent plot for (15)~Eunomia is shown
in Figure \ref{15_vs_lit}. These results appear closer to, but not
quite, Gaussian.

In most cases, the MCMC results are fairly similar to those of the
Nelder-Mead algorithm (Table \ref{neldermead}) with some exceptions,
such as [704;7461] where Nelder-Mead is very close to the literature
value unlike the MCMC algorithm. Conversely, in some other cases such
as [13;14689] the MCMC result is much better.  Overall, the results
for both algorithms are in line with each other.

\section{Conclusions}

We have successfully developed and implemented a new
Adaptive-Metropolis algorithm for asteroid mass estimation.  Our
results agree with previously published mass estimates, but suggest
that the published uncertainties may be misleading as a consequence of
using linearized mass-estimation methods. Future work on the
methodology includes extending the algorithm to use multiple
perturbers and/or test asteroids simultaneously, more robust
observational error and outlier rejection models, and accounting for
systematic errors automatically. Finally, we intend to expand the
application of the method by systematically obtaining mass estimates
using existing astrometry and, eventually, from the Gaia mission once
the data is released.

\section*{Acknowledgements}

We are grateful for the constructive criticism offered by the two
anonymous reviewers which helped to improve the paper. We also thank
Karri Muinonen for helpful discussions and, in particular, for sharing
the basic idea of the marching method with us. This work was supported
by grants \#299543 and \#307157 from the Academy of Finland. This
research has made use of NASA's Astrophysics Data System.

\section*{References}

\bibliographystyle{apa}
\bibliography{gradu.bib}

\end{document}